\documentclass[12pt]{article}
\usepackage{amsmath,amssymb,amsfonts}
\usepackage{graphicx}
\usepackage{subcaption}
\usepackage{multirow}
\usepackage{array}
\usepackage{float}
\usepackage{geometry}
\usepackage{hyperref}
\usepackage{enumitem}
\usepackage{booktabs}
\usepackage[sort&compress, numbers, merge]{natbib}
\hypersetup{
    colorlinks=true,
    linkcolor=blue,
    citecolor=blue,
}
\begin{document}

\begin{titlepage}
\begin{center}

\vspace{3.0cm}
{\large \textbf{Baryon Asymmetry Transfer to Majorana Dark Matter from a Colored Partner}}

\vspace{1.0cm}
{Wei Huang}, 
{Hongjian Du},
{Feng Luo}
and 
{Michihisa Takeuchi} 
\vspace{1.0cm}

{\it
{School of Physics and Astronomy, Sun Yat-sen University, Zhuhai 519082, China}\\
}

\vspace{1cm}
\begin{abstract}
We propose a mechanism in which a primordial baryon asymmetry determines the relic abundance of self-conjugate dark matter through a heavier partner sector. In a simplified model with a Majorana fermion $\chi$ and a colored scalar partner $\tilde q$, QCD annihilations deplete the symmetric partner population while a residual asymmetric component survives as a temporary reservoir. For feeble portal couplings, this asymmetry is transferred to $\chi$ at late times, after dark matter freeze-out. The final relic abundance is controlled by the amount of surviving asymmetry, the timing of its transfer, and the subsequent dark matter annihilation.
Solving the coupled Boltzmann equations with the relevant chemical-potential effects, we find that an intermediate range of portal couplings can reproduce the observed relic abundance without requiring a compressed mass spectrum. The qualitative mechanism remains robust against variations in the mass hierarchy, dark matter annihilation mode, and lepton asymmetry. The small portal couplings also make the colored partner potentially long lived, allowing LHC searches for heavy long-lived particles to constrain part of the viable parameter space.
\end{abstract}
\end{center}

\end{titlepage}

\setcounter{footnote}{0}

\section{Introduction}
\label{sec:intro}

The coexistence of visible matter and dark matter (DM) in the Universe, together with their similar cosmological abundances, remains one of the central puzzles of modern cosmology. 
Astronomical observations indicate that the present-day dark matter density is approximately five times larger than the baryonic matter density~\cite{Planck:2018vyg}. 
The numerical coincidence between the baryon and dark matter abundances has motivated the framework of asymmetric dark matter (ADM), in which the dark matter relic density is related to a primordial particle-antiparticle asymmetry~\cite{Hooper:2004dc, Zurek:2013wia, Kumar:2013vba, Petraki:2013wwa, Kaplan:2009ag, Shelton:2010ta}. 
A variety of mechanisms have been proposed to generate and transfer such an asymmetry between the visible and dark sectors~\cite{Suematsu:2005kp, Cohen:2009fz, Cohen:2010kn, Belyaev:2010kp, An:2009vq, Nardi:2008ix, Heckman:2011sw, Davoudiasl:2010am, Haba:2010bm, Blennow:2010qp, Buckley:2010ui, Frandsen:2011kt, Hall:2010jx, Dutta:2010va, Dutta:2024bnt, Falkowski:2011xh, Borah:2024wos, Iminniyaz:2011yp, Graesser:2011wi, Rashidin:2025zgh}.

In conventional ADM scenarios, the dark matter particle itself carries the conserved particle-antiparticle asymmetry that determines its relic abundance. 
This picture, however, cannot be directly applied to a self-conjugate dark matter candidate, such as a Majorana fermion, for which particle and antiparticle are identical. 
This raises a simple but important question: \emph{can a primordial baryon asymmetry nevertheless determine the relic abundance of self-conjugate dark matter?}

In this work, we propose a mechanism that provides an affirmative answer to this question. 
The central idea is that the asymmetry need not be stored directly in the dark matter particle. 
Instead, a primordial baryon asymmetry can be shared between the Standard Model (SM) quark sector and a heavier partner sector, with the asymmetric component in the latter acting as a temporary reservoir. 
As the Universe evolves, this partner-sector asymmetry can survive the depletion of the symmetric partner population and subsequently be transferred to the SM quark sector through the decay of the partner into dark matter and a quark. 
At the same time, the amount of asymmetry that survives in the partner sector determines the abundance of the self-conjugate dark matter particle. 
This provides a qualitatively different realization of asymmetric dark matter in which a self-conjugate dark matter particle can inherit its relic abundance from a primordial baryon asymmetry shared between the SM quark and partner sectors.

To realize this idea, we consider a simplified model containing a Majorana fermion $\chi$, a complex scalar partner $\tilde q$, and the SM quark flavor $q$ associated with the partner, interacting through
\begin{equation}
\mathcal{L}_{\rm portal}
=
-\lambda\,\tilde q\,\overline{q}\,
\frac{1-\gamma_5}{2}\,\chi
+\text{h.c.}.
\label{eq:portal}
\end{equation}
At sufficiently high temperatures, $T\gg m_{\tilde q}$, we assume that a net baryon asymmetry is distributed among the SM quark flavors and the $\tilde q$ sector. 
The origin of this distribution of the primordial asymmetry is not addressed in this work; it may arise from chemical equilibrium established through additional high-temperature interactions that become ineffective at lower temperatures, or from other dynamics in the dark sector.
Our focus is instead on its subsequent cosmological evolution, in particular on how the asymmetry stored in the $\tilde q$ sector evolves and how its eventual transfer to the SM quark sector is correlated with the relic abundance of $\chi$.

\begin{figure}[t]
\centering
\includegraphics[width=0.7\textwidth]{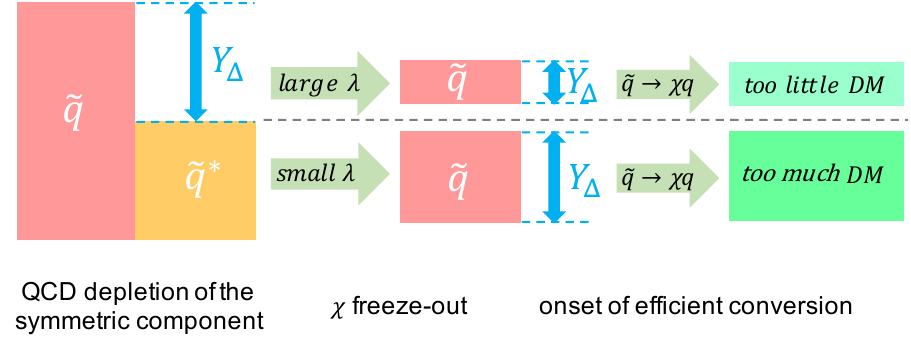}
\caption{
Schematic illustration of the baryon asymmetry transfer mechanism.
The sizes of the colored blocks represent the relative number densities
of the corresponding species.
The primordial baryon asymmetry is shared between the SM quark and
partner sectors, with $Y_\Delta$ denoting the asymmetric component
stored in the partner sector.
The subsequent evolution involves QCD depletion of the symmetric
partner component and late-time transfer of the surviving partner
asymmetry into dark matter through $\tilde q\to\chi q$.
The amount of asymmetry that survives in the partner reservoir depends
on the portal coupling $\lambda$.
}
\label{fig:conceptual_plot}
\end{figure}

The essential dynamics are governed by the interplay between the rates of the
$\tilde q\leftrightarrow\chi$ conversion processes and the Hubble expansion
rate. For the epoch relevant to our mechanism, the total conversion rate is
initially below the Hubble expansion rate, so that the partner and dark matter
sectors are not in chemical equilibrium.
As the Universe cools, the
conversion processes become effective at later times, around and after
the thermal freeze-out of $\chi$, while QCD annihilations deplete the
symmetric component of the $\tilde q$ population. 
The asymmetric component is eventually transferred from $\tilde{q}$ to $\chi$ through 
decay or scattering processes, 
simultaneously contributing to the dark matter relic abundance and
transferring the baryon asymmetry stored in the partner sector to the SM
quark sector, as schematically illustrated in Fig.~\ref{fig:conceptual_plot}.

The amount of asymmetric component that survives in the partner sector
depends sensitively on the portal coupling $\lambda$. For a sufficiently
large $\lambda$, the late-time conversion is more efficient and the
partner asymmetry is more strongly depleted, leaving only a small
asymmetric component to be transferred to dark matter. In contrast, for
a sufficiently small $\lambda$, the surviving asymmetric excess can overproduce
dark matter.
The observed relic abundance is consequently obtained for an
intermediate $\lambda$ that allows a substantial but appropriately sized
partner-sector asymmetry to survive.

For the benchmark mass choice $m_{\tilde q}=2~\mathrm{TeV}$ and $m_\chi=500~\mathrm{GeV}$, we find that the observed dark matter relic abundance can be reproduced for a feeble coupling of order
\begin{equation}
    \lambda \sim 10^{-9}.
\end{equation}
The precise value depends on the mass spectrum and the assumed primordial asymmetries. 
We quantify this cosmological evolution by numerically solving the coupled Boltzmann equations governing $\chi$, $\tilde q$, and $\tilde q^*$.

An important feature of this mechanism is that it does not require the partner and dark matter states to be nearly degenerate. 
In conventional coannihilation scenarios~\cite{Griest:1990kh}, the heavier partner abundance is typically Boltzmann suppressed, and a compressed spectrum, $m_{\tilde q}\simeq m_\chi$, is therefore required for the partner to remain sufficiently abundant to influence the dark matter relic abundance. 
In the present framework, by contrast, the relevant quantity is the asymmetric component that survives the annihilation and conversion history rather than the total thermal abundance of the partner. 
The partner can therefore act as an asymmetric reservoir even when it is significantly heavier than dark matter, allowing the mechanism to operate for hierarchical spectra with $m_{\tilde q}\gg m_\chi$.

Our scenario also differs conceptually from conversion-driven freeze-out~\cite{Garny:2017rxs}. 
Although both frameworks involve feeble interactions between dark matter and a heavier partner, conversion-driven freeze-out is formulated for a symmetric dark sector, where conversion processes control the freeze-out of the dark matter abundance. 
It likewise relies on a compressed mass spectrum, $m_{\tilde q}\simeq m_\chi$, so that the heavier partner remains sufficiently abundant during the non-relativistic evolution to maintain efficient conversion with dark matter. 
In our framework, by contrast, the conversion processes control the survival and eventual transfer of a primordial asymmetry stored in the partner sector. 

The feeble coupling required for efficient asymmetry transfer has an additional phenomenological consequence. 
Since $\tilde q$ is a colored scalar, its small coupling to $\chi$ and $q$ can render it long-lived on detector time scales~\cite{Heisig:2024xbh, Arina:2025zpi}. 
After hadronization, such long-lived colored states form $R$-hadrons and can therefore be searched for through heavy stable charged particle (HSCP) signatures at the LHC~\cite{ATLAS:2019gqq, ATLAS:2022pib}. 
We utilize the ATLAS full Run-2 HSCP search results to constrain the parameter space for both the $\tilde t$ and $\tilde b$ partner scenarios across their viable lifetime ranges. 
Thus, the same feeble coupling that enables the cosmological asymmetry-transfer mechanism also leads to potentially observable long-lived-particle signatures at colliders.

The remainder of this paper is organized as follows.
In Sec.~\ref{sec:model_and_BE}, we introduce the simplified model and derive the coupled Boltzmann equations governing the evolution of the relevant particle abundances and asymmetries.
In Sec.~\ref{sec:chemical_potential}, we discuss the treatment of chemical potentials and their role in closing the Boltzmann equations.
In Sec.~\ref{sec:numerical_analysis}, we present the numerical results and investigate the dependence of the relic abundance on the model parameters.
In Sec.~\ref{sec:collider}, we study the collider constraints from long-lived particle searches at the LHC.
Finally, we summarize our main conclusions in Sec.~\ref{sec:conclusion}.

\section{The Model and Boltzmann Equations}
\label{sec:model_and_BE}

\subsection{Simplified model}
\label{subsec:model}

To quantify the late-time dynamical evolution of the asymmetry transfer, we elaborate on the simplified model structure introduced in Sec.~\ref{sec:intro}. 
Beyond the portal interaction $\mathcal{L}_{\rm portal}$ given in Eq.~(\ref{eq:portal}), the complex scalar partner $\tilde{q}$ is a color-triplet state that couples directly to the SM gauge fields through the gauge-kinetic term
\begin{equation}
\label{eq:gauge}
\mathcal{L}_{\rm gauge} = (D_\mu \tilde{q})^\dagger (D^\mu \tilde{q}) \, .
\end{equation}
In particular, the covariant derivative acting on the color-triplet scalar $\tilde{q}$ is explicitly given by
\begin{equation}
\label{eq:cov_derivative}
D_\mu \tilde{q}_i = \left( \partial_\mu \delta_{ij} + i g_s T^a_{ij} A_\mu^a \right) \tilde{q}_j \, ,
\end{equation}
where $g_s$ denotes the strong coupling constant, $A_\mu^a$ ($a = 1, \dots, 8$) represents the gluon gauge fields, and $T^a$ are the $SU(3)_c$ generators in the fundamental representation, satisfying $\text{Tr}(T^a T^b) = \frac{1}{2}\delta^{ab}$.
These gauge interactions mediate rapid $\tilde{q}\tilde{q}^*$ pair annihilations, dominated by QCD processes, including Sommerfeld enhancement and bound-state effects~\cite{DeSimone:2014qkh, Harigaya:2014dwa, Ellis:2015vaa, An:2016gad, Liew:2016hqo, Harz:2018csl, Fukuda:2018ufg, Keung:2017kot, Becker:2026icc}.
They therefore efficiently deplete the symmetric component of $\tilde{q}$ in the early Universe while leaving a potentially nonzero asymmetric component.

Together with $\mathcal{L}_{\rm portal}$, these interactions follow the simplified model setup studied in Ref.~\cite{Garny:2017rxs}, where $\tilde{q}$ mimics the essential gauge and portal dynamics of a right-handed squark in the Minimal Supersymmetric Standard Model (MSSM), without incorporating the full set of supersymmetric interactions.

Apart from $\mathcal{L}_{\rm portal}$, we do not explicitly specify the additional interactions responsible for maintaining the thermal history of $\chi$.
We assume that $\chi$ is initially in full chemical equilibrium with the thermal plasma and remains in chemical equilibrium until its ordinary thermal freeze-out.
The corresponding number-changing interactions are taken to be sufficiently efficient such that
\begin{equation}
Y_\chi \simeq Y_\chi^{\rm eq}
\end{equation}
at temperatures above the freeze-out temperature.
After freeze-out, these interactions become inefficient, and $Y_\chi$ departs from its equilibrium value, leaving a residual thermal relic abundance.

For the parameter choices considered below, this ordinary thermal relic abundance is assumed to be subdominant to the observed dark matter abundance.
The remaining dark matter abundance is subsequently generated through the transfer of the asymmetry stored in $\tilde{q}$ to the $\chi$ population.
Thus, the mechanism studied here does not replace the ordinary thermal freeze-out of $\chi$, but supplements its residual thermal relic abundance with an additional contribution sourced by the asymmetry in the partner sector.

For a Majorana fermion $\chi$, such a thermal history can be realized through additional annihilation channels that are not explicitly specified in the simplified model.
For example, efficient $p$-wave annihilation through a CP-even Higgs-like scalar funnel can maintain chemical equilibrium at high temperatures and subsequently lead to ordinary thermal freeze-out.
Alternatively, additional dark-sector states may participate in the annihilation processes, as in coannihilation scenarios analogous to slepton-neutralino coannihilation in the MSSM~\cite{Ellis:2012nv, Ellis:2015rya}.
The detailed ultraviolet realization of these additional interactions does not affect the asymmetry-transfer dynamics considered here, and we therefore parameterize their net effect through an effective $\chi\chi$ annihilation cross section.

The portal coupling $\lambda$ is assumed to be sufficiently small that the interactions mediated solely by the portal do not maintain chemical equilibrium between $\chi$ and $\tilde q$ during the relevant late-time evolution.
In particular, the conversion processes induced by $\lambda$ can become inefficient compared with the Hubble expansion rate well before the QCD annihilations of the symmetric $\tilde q$ population cease.
Consequently, the $\chi$ and $\tilde q$ abundances cannot be described by a common chemical-equilibrium condition, and the evolution of their abundances must be followed separately.

\subsection{Cross sections and rates}
\label{subsec:rates}

Starting from the Lagrangians in Eqs.~(\ref{eq:portal}) and~(\ref{eq:gauge}), we use the Mathematica package FeynCalc~\cite{Mertig:1990an} to compute the tree-level spin- and color-averaged squared invariant amplitudes $|\overline{\mathcal{M}}|^2$ for the reactions entering the Boltzmann equations.
The relevant processes and their dependence on the coupling parameters are summarized in Table~\ref{tab:coupling}.
Together, these reactions determine the dynamical evolution of the $\tilde{q}$ asymmetry and its transfer to the $\chi$ sector.

\begin{table}[htbp]
\centering
\caption{Reaction types and their coupling dependences.}
\label{tab:coupling}
\begin{tabular}{lll}
\toprule
Reaction type & Reaction & Coupling dependence \\
\midrule
\multirow{3}{*}{Conversion}
& $\tilde{q} \leftrightarrow \chi q$
& $\lambda^2$ \\
& $\tilde{q}g \leftrightarrow q\chi$
& \multirow{2}{*}{$\lambda^2 g_s^2$} \\
& $\tilde{q}\bar{q} \leftrightarrow \chi g$
& \\
\addlinespace
Coannihilation
& $\tilde{q}\chi \leftrightarrow qg$
& $\lambda^2 g_s^2$ \\
\addlinespace
\multirow{3}{*}{Portal annihilation}
& $\chi\chi \leftrightarrow q\bar{q}$
& \multirow{3}{*}{$\lambda^4$} \\
& $\tilde{q}\tilde{q} \leftrightarrow qq$
& \\
& $\tilde{q}\tilde{q}^* \leftrightarrow \chi\chi$
& \\
\addlinespace
\multirow{2}{*}{QCD annihilation}
& $\tilde{q}\tilde{q}^* \leftrightarrow q\bar{q}$
& \multirow{2}{*}{$g_s^4$} \\
& $\tilde{q}\tilde{q}^* \leftrightarrow gg$
& \\
\bottomrule
\end{tabular}
\end{table}

We classify the reactions according to their roles in the asymmetry-transfer dynamics.
Here, conversion processes are defined as reactions that directly exchange particle number between the $\tilde{q}$ and $\chi$ sectors.
They include both the decay/inverse-decay process $\tilde{q}\leftrightarrow\chi q$ and the $2\to2$ conversion scatterings $\tilde{q}g\leftrightarrow q\chi$ and $\tilde{q}\bar{q}\leftrightarrow\chi g$.
Although these two classes of conversion processes have different coupling dependences, they both change the relative abundances of $\tilde{q}$ and $\chi$.
The decay and inverse-decay rates are proportional to $\lambda^2$, while the conversion scatterings involve one portal and one QCD vertex and therefore scale as $\lambda^2 g_s^2$.
In contrast, the process $\tilde q\chi\leftrightarrow qg$ is treated separately as a coannihilation process.
Although this process also changes both $Y_\chi$ and $Y_{\tilde q}$, it removes a $\tilde q$ and a $\chi$ simultaneously rather than converting one species into the other.
We therefore classify it as coannihilation rather than conversion.

In the parameter region relevant for the asymmetry-transfer mechanism,
the portal coupling $\lambda$ is much smaller than the strong coupling
$g_s$, with $g_s$ of order unity.
Consequently, the portal-mediated processes are parametrically suppressed
relative to the QCD interactions.
In particular, the small portal coupling prevents $\tilde{q}$ and $\chi$
from remaining in chemical equilibrium as the Universe cools.
Instead, the two sectors can depart from chemical equilibrium, allowing
the pre-existing asymmetry stored in $\tilde{q}$ to be transferred to
$\chi$ through out-of-equilibrium conversion processes.

The coupling dependences summarized in Table~\ref{tab:coupling} allow us to simplify the coupled Boltzmann equations.
The strong-interaction processes\footnote{For $\tilde{q}\tilde{q}^* \leftrightarrow q\bar{q}$, we note that due to the $s$-channel gluon-mediated diagram, the final-state quark and antiquark are not necessarily of the flavor associated with $\tilde{q}$. Also, although a $t$-channel $\chi$-exchange diagram exists for this process, its contribution to $|\overline{\mathcal{M}}|^2$ is suppressed and negligible.}
$\tilde{q}\tilde{q}^* \leftrightarrow q\bar{q}$ and $\tilde{q}\tilde{q}^* \leftrightarrow gg$ depend solely on $g_s$ and therefore provide rapid pair-annihilation processes that are independent of $\lambda$.
They efficiently deplete the symmetric component of $\tilde{q}$ while leaving its asymmetry unaffected.

By contrast, the portal annihilation processes proportional to $\lambda^4$, namely $\tilde{q}\tilde{q}\leftrightarrow qq$, $\tilde{q}\tilde{q}^*\leftrightarrow\chi\chi$, and the portal-induced contribution to $\chi\chi\leftrightarrow q\bar{q}$, are strongly suppressed relative to the $g_s^4$ and $\lambda^2 g_s^2$ processes for $\lambda\ll g_s$.
We therefore neglect these portal-induced processes in the coupled Boltzmann equations governing the $\tilde{q}$--$\chi$ asymmetry-transfer dynamics.
This approximation does not mean that $\chi\chi$ annihilation is neglected altogether.
As discussed in Sec.~\ref{subsec:model}, the ordinary thermal freeze-out of $\chi$ is assumed to be controlled by additional interactions beyond the simplified model, whose net effect is parameterized by an effective annihilation cross section $\langle\sigma v\rangle_{\chi\chi}$, which determines the residual thermal relic abundance of $\chi$, while the portal-mediated processes considered here govern the subsequent transfer of the $\tilde{q}$ asymmetry.

In addition to the processes listed in Table~\ref{tab:coupling}, electroweak processes are allowed by Eq.~(\ref{eq:gauge}) and by the SM interaction Lagrangian, such as $\tilde{q}\tilde{q}^* \leftrightarrow ZH$ and $\tilde{q}\gamma\leftrightarrow q\chi$.
These electroweak channels have been shown to be subdominant compared to the strong-interaction processes~\cite{Garny:2017rxs}.
Moreover, as will be demonstrated in Sec.~\ref{sec:numerical_analysis}, the relevant asymmetry-transfer dynamics is predominantly controlled by $\lambda$, with the $\lambda^2$ and $\lambda^2g_s^2$ processes governing the relevant portal-induced rates.
Including the electroweak interactions would therefore introduce only a minor numerical correction to the extracted value of $\lambda$, and we do not include them in our Boltzmann system.

All cross sections (multiplied by the corresponding relative velocity of the incoming particles in each process) entering the Boltzmann equations are thermally averaged, denoted by the bracket $\langle\cdot\rangle$, using the standard formula~\cite{Edsjo:1997bg}
\begin{equation}
\langle \sigma_{ij} v \rangle n_i^{\text{eq}} n_j^{\text{eq}}
=
T \frac{g_i g_j}{256 \pi^5}
\int
\frac{p_{ij} p_{ab}}{\sqrt{s}}
|\overline{\mathcal{M}}|^2
K_1\left(\frac{\sqrt{s}}{T}\right)
\,\mathrm{d}s\,\mathrm{d}\cos\theta \,,
\label{eq:thermal_average_cross_section}
\end{equation}
where $K_1$ denotes the modified Bessel function of the second kind of order one.
Here, $g_i$ ($g_j$) represents the internal degrees of freedom of species $i$ ($j$), while $p_{ij}$ ($p_{ab}$) denotes the magnitude of the three-momentum of the initial (final) state particles in the center-of-mass frame.

The thermally averaged decay rate
$\langle\Gamma\rangle_{\tilde{q}\to\chi q}$
is related to the decay width
$\Gamma_{\tilde{q}\to\chi q}$
evaluated in the rest frame of $\tilde{q}$ by
\begin{equation}
\left\langle \Gamma \right\rangle_{\tilde{q} \to \chi q}
=
\Gamma_{\tilde{q} \to \chi q}
\frac{K_1(m_{\tilde{q}}/T)}
{K_2(m_{\tilde{q}}/T)} \,,
\label{eq:thermally_decay_rate}
\end{equation}
where $K_2$ is the modified Bessel function of the second kind of order two.

\subsection{Boltzmann equations}
\label{subsec:boltzmann_equations}

In standard coannihilation scenarios with sufficiently rapid
inter-species conversion, chemical equilibrium among dark-sector particles
is maintained until decoupling, allowing the relic abundance to be described
by a single Boltzmann equation for the total abundance~\cite{Griest:1990kh}.
In our framework, however, the portal coupling $\lambda$ is sufficiently
small that the conversion rates between $\tilde{q}$ and $\chi$ can fall
below the Hubble expansion rate $H(T)$, preventing $\tilde{q}$ and $\chi$
from remaining in chemical equilibrium throughout their subsequent
evolution.

To accurately track this departure from chemical equilibrium, we solve the
coupled Boltzmann equations for $Y_\chi$, $Y_{\tilde q}$, and $Y_{\tilde q^*}$.
This treatment allows the ordinary thermal freeze-out of $\chi$, the QCD
depletion of the symmetric component of $\tilde q$, and the subsequent
transfer of the $\tilde q$ asymmetry to $\chi$ to be followed consistently
within a single framework.

It is convenient to work with the number densities of the dark matter
particle $\chi$, the complex scalar $\tilde q$, and its antiparticle
$\tilde q^*$, normalized by the entropy density $s$:
\begin{equation}
Y_\chi \equiv \frac{n_\chi}{s},
\qquad
Y_{\tilde q} \equiv \frac{n_{\tilde q}}{s},
\qquad
Y_{\tilde q^*} \equiv \frac{n_{\tilde q^*}}{s}.
\end{equation}

To directly focus on the physically relevant quantities---namely, the total
partner abundance
\begin{equation}
Y_\Sigma \equiv Y_{\tilde q} + Y_{\tilde q^*}
\end{equation}
and the net asymmetry
\begin{equation}
Y_\Delta \equiv Y_{\tilde q} - Y_{\tilde q^*},
\end{equation}
we transform the system of Boltzmann equations
into these sum and difference variables.
The coupled system for $Y_\chi$, $Y_\Sigma$, and $Y_\Delta$ is then
\begin{equation}
\begin{aligned}
\frac{\mathrm{d}Y_{\chi}}{\mathrm{d}x}
={}&
-\frac{xs}{H(m_{\chi})}
\left(
1+\frac{T}{3g_{\ast s}}
\frac{\mathrm{d}g_{\ast s}}{\mathrm{d}T}
\right)
\\
&\times
\Biggl\{
\left\langle \sigma v \right\rangle_{\chi\chi}
\left[
Y_{\chi}^2-(Y_{\chi}^{\mathrm{eq}})^2
\right]
\\
&\qquad
+\left\langle \sigma v \right\rangle_{\tilde q\chi\to qg}
\left[
Y_\Sigma Y_\chi
-
Y_\chi^{\mathrm{eq}}
\left(
Y_{\tilde q,\mathrm{w}}^{\mathrm{eq}}
+
Y_{\tilde q^*,\mathrm{w}}^{\mathrm{eq}}
\right)
\right]
\\
&\qquad
-\frac{1}{s}
\left\langle \Gamma \right\rangle_{\tilde q\to\chi}
\left(
\frac{Y_\Sigma+Y_\Delta}{2}
-
Y_\chi
\frac{Y_{\tilde q,\mathrm{w}}^{\mathrm{eq}}}
{Y_\chi^{\mathrm{eq}}}
\right)
\\
&\qquad
-\frac{1}{s}
\left\langle \Gamma \right\rangle_{\tilde q^*\to\chi}
\left(
\frac{Y_\Sigma-Y_\Delta}{2}
-
Y_\chi
\frac{Y_{\tilde q^*,\mathrm{w}}^{\mathrm{eq}}}
{Y_\chi^{\mathrm{eq}}}
\right)
\Biggr\},
\end{aligned}
\label{eq:BE_chi}
\end{equation}

\begin{equation}
\begin{aligned}
\frac{\mathrm{d}Y_\Sigma}{\mathrm{d}x}
={}&
-\frac{xs}{H(m_{\chi})}
\left(
1+\frac{T}{3g_{\ast s}}
\frac{\mathrm{d}g_{\ast s}}{\mathrm{d}T}
\right)
\\
&\times
\Biggl\{
\frac{1}{2}
\left\langle \sigma v \right\rangle_{\tilde q\tilde q^*}
\Bigl[
(Y_\Sigma+Y_\Delta)(Y_\Sigma-Y_\Delta)
-4Y_{\tilde q,\mathrm{w/o}}^{\mathrm{eq}}
Y_{\tilde q^*,\mathrm{w/o}}^{\mathrm{eq}}
\Bigr]
\\
&\qquad
+\left\langle \sigma v \right\rangle_{\tilde q\chi\to qg}
\left[
Y_\chi Y_\Sigma
-
Y_\chi^{\mathrm{eq}}
\left(
Y_{\tilde q,\mathrm{w}}^{\mathrm{eq}}
+
Y_{\tilde q^*,\mathrm{w}}^{\mathrm{eq}}
\right)
\right]
\\
&\qquad
+\frac{1}{s}
\left\langle \Gamma \right\rangle_{\tilde q\to\chi}
\left(
\frac{Y_\Sigma+Y_\Delta}{2}
-
Y_\chi
\frac{Y_{\tilde q,\mathrm{w}}^{\mathrm{eq}}}
{Y_\chi^{\mathrm{eq}}}
\right)
\\
&\qquad
+\frac{1}{s}
\left\langle \Gamma \right\rangle_{\tilde q^*\to\chi}
\left(
\frac{Y_\Sigma-Y_\Delta}{2}
-
Y_\chi
\frac{Y_{\tilde q^*,\mathrm{w}}^{\mathrm{eq}}}
{Y_\chi^{\mathrm{eq}}}
\right)
\Biggr\},
\end{aligned}
\label{eq:BE_Sigma}
\end{equation}

\begin{equation}
\begin{aligned}
\frac{\mathrm{d}Y_\Delta}{\mathrm{d}x}
={}&
-\frac{xs}{H(m_{\chi})}
\left(
1+\frac{T}{3g_{\ast s}}
\frac{\mathrm{d}g_{\ast s}}{\mathrm{d}T}
\right)
\\
&\times
\Biggl\{
\left\langle \sigma v \right\rangle_{\tilde q\chi\to qg}
\left[
Y_\chi Y_\Delta
-
Y_\chi^{\mathrm{eq}}
\left(
Y_{\tilde q,\mathrm{w}}^{\mathrm{eq}}
-
Y_{\tilde q^*,\mathrm{w}}^{\mathrm{eq}}
\right)
\right]
\\
&\qquad
+\frac{1}{s}
\left\langle \Gamma \right\rangle_{\tilde q\to\chi}
\left(
\frac{Y_\Sigma+Y_\Delta}{2}
-
Y_\chi
\frac{Y_{\tilde q,\mathrm{w}}^{\mathrm{eq}}}
{Y_\chi^{\mathrm{eq}}}
\right)
\\
&\qquad
-\frac{1}{s}
\left\langle \Gamma \right\rangle_{\tilde q^*\to\chi}
\left(
\frac{Y_\Sigma-Y_\Delta}{2}
-
Y_\chi
\frac{Y_{\tilde q^*,\mathrm{w}}^{\mathrm{eq}}}
{Y_\chi^{\mathrm{eq}}}
\right)
\Biggr\}.
\end{aligned}
\label{eq:BE_Delta}
\end{equation}

Here,
$
x \equiv m_\chi/T
$
is used as the time-evolution variable.
The entropy density $s$ and the rescaled Hubble parameter $H(m_\chi)$ are
defined as
\begin{equation}
s
=
\frac{2\pi^2}{45}g_{\ast s}T^3
=
\frac{2\pi^2}{45}g_{\ast s}
\frac{m_\chi^3}{x^3},
\qquad
H(m_\chi)
\equiv H(T)x^2
=
\sqrt{\frac{4\pi^3g_\ast}{45}}
\frac{m_\chi^2}{m_{\rm pl}},
\end{equation}
where $m_{\rm pl}\simeq1.22\times10^{19}$~GeV is the Planck mass, and
$g_{\ast s}$ and $g_\ast$ denote the effective relativistic degrees of
freedom for entropy and energy density, respectively~\cite{Saikawa:2018rcs}.

The physical meaning of the chemical-potential-dependent equilibrium
abundances $Y_{\tilde q,\mathrm{w}}^{\mathrm{eq}}$ and
$Y_{\tilde q^*,\mathrm{w}}^{\mathrm{eq}}$, as well as the generalized rates
$\langle\Gamma\rangle_{\tilde q\to\chi}$ and
$\langle\Gamma\rangle_{\tilde q^*\to\chi}$ appearing in
Eqs.~(\ref{eq:BE_chi})--(\ref{eq:BE_Delta}), can be understood from the
corresponding detailed-balance relations.

The subscript ``$\mathrm{w/o}$'' denotes the thermal equilibrium state
evaluated at vanishing chemical potential, $\mu=0$.
Under the Maxwell--Boltzmann (MB) approximation\footnote{%
MB statistics is adopted here to simplify the evaluation of collision
integrals in the Boltzmann equations, consistent with the derivation of
Eqs.~(\ref{eq:thermal_average_cross_section}) and
(\ref{eq:thermally_decay_rate}).
In the absence of Bose condensation or Fermi degeneracy, MB
statistics provides a good approximation and is commonly used in the
treatment of Boltzmann equations~\cite{Kolb:1990vq,Gondolo:1990dk}.
It also allows the chemical-potential dependence
$e^{\pm\mu_q/T}$ to be factored exactly out of the phase-space integrals,
leading to a simple detailed-balance relation.}
the equilibrium number densities of quarks and scalar partners at
$\mu=0$ are
\begin{equation}
n_{q,\mathrm{w/o}}^{\mathrm{eq}}
=
n_{\bar q,\mathrm{w/o}}^{\mathrm{eq}}
=
\frac{g_q}{2\pi^2}
\int_{m_q}^{\infty}
E\sqrt{E^2-m_q^2}
\exp\left(-\frac{E}{T}\right)
\,\mathrm{d}E
=
\frac{g_q}{2\pi^2}
Tm_q^2
K_2\left(\frac{m_q}{T}\right),
\end{equation}
and
\begin{equation}
n_{\tilde q,\mathrm{w/o}}^{\mathrm{eq}}
=
n_{\tilde q^*,\mathrm{w/o}}^{\mathrm{eq}}
=
\frac{g_{\tilde q}}{2\pi^2}
\int_{m_{\tilde q}}^{\infty}
E\sqrt{E^2-m_{\tilde q}^2}
\exp\left(-\frac{E}{T}\right)
\,\mathrm{d}E
=
\frac{g_{\tilde q}}{2\pi^2}
Tm_{\tilde q}^2
K_2\left(\frac{m_{\tilde q}}{T}\right),
\label{eq:n_tildeq_eq_wo}
\end{equation}
where $g_q=6$ and $g_{\tilde q}=3$ denote the internal degrees of freedom
of $q$ and $\tilde q$, respectively.
We note that this modified Bessel-function representation is valid for
arbitrary $m_i/T$ and smoothly interpolates between the relativistic and
non-relativistic regimes.

Assuming that all SM particles remain in local thermal equilibrium, the
number densities of quarks and antiquarks in the presence of the quark
chemical potential $\mu_q$ satisfy
\begin{equation}
n_q
=
n_{q,\mathrm{w/o}}^{\mathrm{eq}}
e^{+\mu_q/T},
\qquad
n_{\bar q}
=
n_{\bar q,\mathrm{w/o}}^{\mathrm{eq}}
e^{-\mu_q/T}.
\end{equation}
Here $\mu_q$ denotes the chemical potential of the quark flavor associated
with $\tilde q$, as will be discussed in detail in
Sec.~\ref{sec:chemical_potential}.

Detailed balance then allows the collision terms governing the conversion
processes to be written in a common form.
For the scattering process
$\tilde q g\leftrightarrow q\chi$, we have
\begin{equation}
\langle\sigma v\rangle_{q\chi\to\tilde q g}
n_\chi^{\mathrm{eq}}
n_{q,\mathrm{w/o}}^{\mathrm{eq}}
=
\langle\sigma v\rangle_{\tilde q g\to q\chi}
n_{\tilde q,\mathrm{w/o}}^{\mathrm{eq}}
n_g^{\mathrm{eq}}.
\label{eq:detailed_balance_example}
\end{equation}
Defining
\begin{equation}
\langle\Gamma\rangle_{\tilde q g\to q\chi}
\equiv
\langle\sigma v\rangle_{\tilde q g\to q\chi}
n_g^{\mathrm{eq}},
\end{equation}
the corresponding collision term becomes
\begin{equation}
\begin{aligned}
&
\langle\sigma v\rangle_{\tilde q g\to q\chi}
n_{\tilde q}n_g
-
\langle\sigma v\rangle_{q\chi\to\tilde q g}
n_qn_\chi
\\
={}&
\langle\Gamma\rangle_{\tilde q g\to q\chi}
\left(
n_{\tilde q}
-
n_\chi
\frac{n_{\tilde q,\mathrm{w}}^{\mathrm{eq}}}
{n_\chi^{\mathrm{eq}}}
\right),
\end{aligned}
\end{equation}
where $n_g=n_g^{\mathrm{eq}}$ is also used.
The effective equilibrium density $n_{\tilde q,\mathrm{w}}^{\mathrm{eq}}$ is
defined as
\begin{equation}
n_{\tilde q,\mathrm{w}}^{\mathrm{eq}}
\equiv
n_{\tilde q,\mathrm{w/o}}^{\mathrm{eq}}
e^{+\mu_q/T}.
\end{equation}
We emphasize that this quantity does not correspond to an independent
chemical potential assigned to $\tilde q$, nor does it imply chemical
equilibrium between $\tilde q$ and $\chi$.
Rather, it is an effective equilibrium density induced by the chemical
potential of the SM quark bath and enters the detailed-balance relation
for the portal conversion processes.
In Sec.~\ref{sec:chemical_potential}, $\mu_q/T$ will be expressed in terms
of the net yield $Y_\Delta$, thereby closing the Boltzmann system
self-consistently.

In terms of the yield $Y=n/s$, we correspondingly define
\begin{equation}
Y_{\tilde q,\mathrm{w}}^{\mathrm{eq}}
\equiv
Y_{\tilde q,\mathrm{w/o}}^{\mathrm{eq}}
e^{+\mu_q/T},
\qquad
Y_{\tilde q,\mathrm{w/o}}^{\mathrm{eq}}
\equiv
\frac{n_{\tilde q,\mathrm{w/o}}^{\mathrm{eq}}}{s}.
\label{eq:y_tildeq_eq_wo}
\end{equation}

Similarly, for the decay and inverse-decay process
$\tilde q\leftrightarrow\chi q$, detailed balance gives
\begin{equation}
\langle\Gamma\rangle_{\tilde q\to\chi q}
n_{\tilde q,\mathrm{w/o}}^{\mathrm{eq}}
=
\langle\sigma v\rangle_{\chi q\to\tilde q}
n_\chi^{\mathrm{eq}}
n_{q,\mathrm{w/o}}^{\mathrm{eq}}.
\label{eq:detailed_balance_decay}
\end{equation}
The corresponding net interaction term can therefore be written as
\begin{equation}
\begin{aligned}
&
\langle\Gamma\rangle_{\tilde q\to\chi q}n_{\tilde q}
-
\langle\sigma v\rangle_{\chi q\to\tilde q}
n_\chi n_q
\\
={}&
\langle\Gamma\rangle_{\tilde q\to\chi q}
\left(
n_{\tilde q}
-
n_\chi
\frac{n_{\tilde q,\mathrm{w}}^{\mathrm{eq}}}
{n_\chi^{\mathrm{eq}}}
\right).
\end{aligned}
\end{equation}
For decays of $\tilde q$ into $\chi$ and additional SM particles\footnote{%
For example, when $\tilde q$ is a $\tilde t$, the conversion may proceed
through three-body decay $\tilde t\leftrightarrow\chi b W^+$
and four-body decays such as
$\tilde t\leftrightarrow\chi b\bar{\tau}\nu_\tau$.
}
the corresponding collision terms can be incorporated in the same
detailed-balance form, provided the relevant SM chemical potentials are
included consistently.
We therefore use the notation
\begin{equation}
\langle\Gamma\rangle_{\tilde q\to\chi\mathrm{SM}}
\left(
n_{\tilde q}
-
n_\chi
\frac{n_{\tilde q,\mathrm{w}}^{\mathrm{eq}}}
{n_\chi^{\mathrm{eq}}}
\right)
\end{equation}
for the net decay/inverse-decay contribution.

For the scattering process
$\tilde q\bar q \leftrightarrow g\chi$, detailed balance requires
\begin{equation}
\langle\sigma v\rangle_{\tilde q\bar q\to g\chi}
n_{\tilde q,\mathrm{w/o}}^{\mathrm{eq}}
n_{\bar q,\mathrm{w/o}}^{\mathrm{eq}}
=
\langle\sigma v\rangle_{g\chi\to\tilde q\bar q}
n_g^{\mathrm{eq}}n_\chi^{\mathrm{eq}}.
\end{equation}
Defining
\begin{equation}
\langle\Gamma\rangle_{\tilde q\bar q\to g\chi}
\equiv
\langle\sigma v\rangle_{\tilde q\bar q\to g\chi}
n_{\bar q,\mathrm{w/o}}^{\mathrm{eq}},
\end{equation}
the collision term becomes
\begin{equation}
\begin{aligned}
&
\langle\sigma v\rangle_{\tilde q\bar q\to g\chi}
n_{\tilde q}n_{\bar q}
-
\langle\sigma v\rangle_{g\chi\to\tilde q\bar q}
n_gn_\chi
\\
={}&
\langle\Gamma\rangle_{\tilde q\bar q\to g\chi}
e^{-\mu_q/T}
\left(
n_{\tilde q}
-
n_\chi
\frac{n_{\tilde q,\mathrm{w}}^{\mathrm{eq}}}
{n_\chi^{\mathrm{eq}}}
\right).
\end{aligned}
\end{equation}

For the coannihilation process
$\tilde q\chi\leftrightarrow qg$, detailed balance gives
\begin{equation}
\langle\sigma v\rangle_{\tilde q\chi\to qg}
n_{\tilde q,\mathrm{w/o}}^{\mathrm{eq}}
n_\chi^{\mathrm{eq}}
=
\langle\sigma v\rangle_{qg\to\tilde q\chi}
n_{q,\mathrm{w/o}}^{\mathrm{eq}}
n_g^{\mathrm{eq}},
\end{equation}
which leads to
\begin{equation}
\begin{aligned}
&
\langle\sigma v\rangle_{\tilde q\chi\to qg}
n_{\tilde q}n_\chi
-
\langle\sigma v\rangle_{qg\to\tilde q\chi}
n_qn_g
\\
={}&
\langle\sigma v\rangle_{\tilde q\chi\to qg}
\left(
n_{\tilde q}n_\chi
-
n_{\tilde q,\mathrm{w}}^{\mathrm{eq}}
n_\chi^{\mathrm{eq}}
\right).
\end{aligned}
\end{equation}

The corresponding relations for the CP-conjugate processes follow
analogously.
In particular,
\begin{equation}
n_{\tilde q^*,\mathrm{w}}^{\mathrm{eq}}
\equiv
n_{\tilde q^*,\mathrm{w/o}}^{\mathrm{eq}}
e^{-\mu_q/T},
\end{equation}
and hence
\begin{equation}
Y_{\tilde q^*,\mathrm{w}}^{\mathrm{eq}}
\equiv
Y_{\tilde q^*,\mathrm{w/o}}^{\mathrm{eq}}
e^{-\mu_q/T}.
\end{equation}
Since
\begin{equation}
Y_{\tilde q^*,\mathrm{w/o}}^{\mathrm{eq}}
=
Y_{\tilde q,\mathrm{w/o}}^{\mathrm{eq}},
\end{equation}
the two effective equilibrium abundances differ only through the
sign of the quark chemical potential.
By CP invariance, the corresponding interaction rates satisfy
\begin{equation}
\begin{aligned}
\langle\Gamma\rangle_{\tilde{q}^* g \to \bar{q} \chi}
&=
\langle\Gamma\rangle_{\tilde{q} g \to q \chi},
\\
\langle\Gamma\rangle_{\tilde q^*\to\chi\bar q}
&=
\langle\Gamma\rangle_{\tilde q\to\chi q},
\\
\langle\Gamma\rangle_{\tilde q^*q \to g\chi}
&=
\langle\Gamma\rangle_{\tilde q\bar q \to g\chi},
\\
\langle\sigma v\rangle_{\tilde q^*\chi\to\bar qg}
&=
\langle\sigma v\rangle_{\tilde q\chi\to qg}.
\end{aligned}
\end{equation}

Combining the decay and scattering contributions, we define the generalized
conversion rates
\begin{equation}
\begin{aligned}
\langle\Gamma\rangle_{\tilde q\to\chi}
&\equiv
\langle\Gamma\rangle_{\tilde q\to\chi\mathrm{SM}}
+
\langle\Gamma\rangle_{\tilde q g\to q\chi}
+
\langle\Gamma\rangle_{\tilde q\bar q\to g\chi}
e^{-\mu_q/T},
\\
\langle\Gamma\rangle_{\tilde q^*\to\chi}
&\equiv
\langle\Gamma\rangle_{\tilde q\to\chi\mathrm{SM}}
+
\langle\Gamma\rangle_{\tilde q g\to q\chi}
+
\langle\Gamma\rangle_{\tilde q\bar q\to g\chi}
e^{+\mu_q/T}.
\end{aligned}
\end{equation}

For the $\tilde q\tilde q^*$ pair-annihilation processes, detailed balance
requires
\begin{equation}
\langle\sigma v\rangle_{\tilde q\tilde q^*\to X\bar X}
n_{\tilde q,\mathrm{w/o}}^{\mathrm{eq}}
n_{\tilde q^*,\mathrm{w/o}}^{\mathrm{eq}}
=
\langle\sigma v\rangle_{X\bar X\to\tilde q\tilde q^*}
n_{X,\mathrm{w/o}}^{\mathrm{eq}}
n_{\bar X,\mathrm{w/o}}^{\mathrm{eq}},
\end{equation}
where $X\bar X$ denotes either $gg$ or $q\bar q$.
The corresponding collision term can therefore be written as
\begin{equation}
\begin{aligned}
&
\langle\sigma v\rangle_{\tilde q\tilde q^*\to X\bar X}
n_{\tilde q}n_{\tilde q^*}
-
\langle\sigma v\rangle_{X\bar X\to\tilde q\tilde q^*}
n_Xn_{\bar X}
\\
={}&
\langle\sigma v\rangle_{\tilde q\tilde q^*\to X\bar X}
\left[
n_{\tilde q}n_{\tilde q^*}
-
n_{\tilde q,\mathrm{w/o}}^{\mathrm{eq}}
n_{\tilde q^*,\mathrm{w/o}}^{\mathrm{eq}}
\frac{n_Xn_{\bar X}}
{n_{X,\mathrm{w/o}}^{\mathrm{eq}}
n_{\bar X,\mathrm{w/o}}^{\mathrm{eq}}}
\right]
\\
={}&
\langle\sigma v\rangle_{\tilde q\tilde q^*\to X\bar X}
\left[
n_{\tilde q}n_{\tilde q^*}
-
n_{\tilde q,\mathrm{w/o}}^{\mathrm{eq}}
n_{\tilde q^*,\mathrm{w/o}}^{\mathrm{eq}}
\right].
\end{aligned}
\end{equation}
In the last step, we have used the fact that the SM bath remains in local
thermal equilibrium.
For a particle-antiparticle pair, the chemical-potential factors cancel,
so that
\begin{equation}
\frac{n_Xn_{\bar X}}
{n_{X,\mathrm{w/o}}^{\mathrm{eq}}
n_{\bar X,\mathrm{w/o}}^{\mathrm{eq}}}
=1.
\end{equation}
Thus, the QCD pair-annihilation collision term is independent of the SM
chemical potential. It contributes equally to the evolution of
$Y_{\tilde q}$ and $Y_{\tilde q^*}$ and therefore cancels in their
difference, leaving the net asymmetry $Y_\Delta$ unchanged.
It consequently affects only the total partner abundance $Y_\Sigma$.

In addition to the tree-level
$\tilde q\tilde q^*\to gg$ and
$\tilde q\tilde q^*\to q\bar q$ processes, which provide the dominant
$2\to2$ pair-annihilation channels, we include non-perturbative Sommerfeld
enhancement and bound-state effects.
These effects are incorporated into the effective
$\tilde q\tilde q^*$ pair-annihilation cross section
$\langle\sigma v\rangle_{\tilde q\tilde q^*}$ following the method described
in Ref.~\cite{Ellis:2015vaa}.

Finally, we specify the effective thermally averaged annihilation cross
section $\langle\sigma v\rangle_{\chi\chi}$ governing the ordinary thermal
freeze-out of $\chi$.
The portal interaction induces the $t$-channel process
$\chi\chi\to q\bar q$, whose contribution is proportional to $\lambda^4$
and is negligible for the parameter range considered here.
The ordinary thermal freeze-out of $\chi$ is instead assumed to be governed
by the additional interactions discussed in Sec.~\ref{subsec:model}.
We therefore treat their net effect phenomenologically through
$\langle\sigma v\rangle_{\chi\chi}$.

We parameterize this effective annihilation cross section using the standard
partial-wave expansion,
\begin{equation}
\langle\sigma v\rangle_{\chi\chi}
=
C_s
+
C_p\frac{T}{m_\chi},
\label{eq:chichi_s_p_wave}
\end{equation}
where $C_s$ and $C_p$ denote the $s$-wave and $p$-wave coefficients,
respectively.

In the numerical analysis presented in
Sec.~\ref{sec:numerical_analysis}, we consider two representative choices:
an $s$-wave-dominated case,
\begin{equation}
C_s=6\times10^{-9}\ {\rm GeV}^{-2},
\qquad
C_p=0,
\end{equation}
and a $p$-wave-dominated case,
\begin{equation}
C_s=0,
\qquad
C_p=3\times10^{-7}\ {\rm GeV}^{-2}.
\end{equation}
Both choices are selected such that, in the absence of asymmetry transfer
from $\tilde q$, the ordinary thermal freeze-out contribution to the relic
density is
\begin{equation}
    \Omega_\chi^{\rm th} h^2\simeq0.04.
\end{equation}
Thus, the ordinary thermal relic provides a controlled subdominant
contribution to the dark matter abundance, while the remaining abundance
is generated through the late-time transfer of the
$\tilde q$ asymmetry to $\chi$.

\section{Chemical Potentials in the Boltzmann Equations}
\label{sec:chemical_potential}

The cosmic baryon asymmetry is conventionally quantified by the
baryon-to-entropy ratio
\begin{equation}
b \equiv \frac{n_b-n_{\bar b}}{s},
\end{equation}
with the observed value
$b \simeq 9\times10^{-11}$~\cite{Planck:2018vyg}.
In our model, the complex scalar $\tilde q$ carries a nonzero baryon number
shared with SM quarks. A self-consistent treatment of the chemical
potentials is therefore required when describing the coupled evolution of
$\tilde q$, $\tilde q^*$, and $\chi$.
We do not address the dynamical origin of the asymmetry, but instead track
its thermal redistribution among the relevant particle species as the
universe evolves, following the framework of Ref.~\cite{Schwarz:2009ii}.

We define the net number density of a particle species $i$ as
\begin{equation}
\Delta n_i \equiv n_i-n_{\bar i}.
\end{equation}
Throughout this section, the number densities of the SM species are
taken to be their local thermal equilibrium values, characterized by the
corresponding chemical potentials.
The exception is the scalar partner $\tilde q$, whose particle and
antiparticle abundances are generally out of equilibrium.
Its net density,
\begin{equation}
\Delta n_{\tilde q}(T)
\equiv
n_{\tilde q}(T)-n_{\tilde q^*}(T),
\end{equation}
is therefore taken directly from the Boltzmann evolution and is not
replaced by an equilibrium expression.

For species in local thermal equilibrium, we use the exact Fermi--Dirac
and Bose--Einstein distributions when evaluating the net number densities.
The net density is given by
\begin{equation}
\Delta n_i(T,\mu_i)
=
\frac{g_i}{2\pi^2}
\int_{m_i}^{\infty}
E\sqrt{E^2-m_i^2}
\left[
\frac{1}{e^{(E-\mu_i)/T}\pm1}
-
\frac{1}{e^{(E+\mu_i)/T}\pm1}
\right]
\,\mathrm{d}E ,
\label{eq:net_density_exact}
\end{equation}
where the upper sign applies to fermions and the lower sign to bosons.

Since the conserved asymmetries in the early universe are small, the
chemical potentials of the SM species remain in the regime
$\mu_i/T\ll1$.
We may therefore expand Eq.~(\ref{eq:net_density_exact}) to first order in
$\mu_i/T$,
\begin{equation}
\Delta n_i(T,\mu_i)
\simeq
\chi_i(T)\frac{\mu_i}{T},
\label{eq:net_density_linear}
\end{equation}
where
\begin{equation}
\chi_i(T)
\equiv
\left.
\frac{\partial\Delta n_i}
{\partial(\mu_i/T)}
\right|_{\mu_i=0}
=
\frac{g_i}{\pi^2}
\int_{m_i}^{\infty}
\frac{
E\sqrt{E^2-m_i^2}\,e^{E/T}
}{
\left(e^{E/T}\pm1\right)^2
}
\,\mathrm{d}E .
\label{eq:density_coeff}
\end{equation}

This linear expansion is particularly useful for the numerical treatment.
Once the chemical-equilibrium and conservation conditions are imposed,
the ratios $\mu_i/T$ can be expressed algebraically in terms of the
corresponding conserved asymmetries. The chemical-potential factors
$e^{\pm\mu_i/T}$ entering the collision terms of the Boltzmann equations
can then be evaluated directly at each temperature, avoiding the need to
solve an additional implicit system at every integration step.

It is important to distinguish the treatment of quantum statistics in this
section from that adopted in Secs.~\ref{subsec:rates} and
\ref{subsec:boltzmann_equations}.
Here, the full Fermi--Dirac and Bose--Einstein distributions are retained
when evaluating the coefficients $\chi_i(T)$, since the relation between
particle asymmetries and their chemical potentials depends on the quantum
statistics of the relevant species and on their mass-to-temperature ratios.
In contrast, Maxwell--Boltzmann statistics is used for the thermally
averaged reaction rates and collision integrals in Secs.~\ref{subsec:rates}
and \ref{subsec:boltzmann_equations}.
The MB approximation allows the factors $e^{\pm\mu_i/T}$ to be factorized
from the phase-space integrals, leading to a convenient form of the
detailed-balance relations.

The use of different statistical treatments in these two contexts is
justified by their different roles.
In the Boltzmann equations, the MB approximation is used to simplify the
evaluation of the reaction rates and collision terms, whereas the
coefficients $\chi_i(T)$ introduced in this section are used to determine
the relation between the physical particle asymmetries and the corresponding
chemical potentials.
Since the quantum-statistical corrections to the relevant reaction rates
are expected to be small in the parameter range considered here, the MB
treatment provides a sufficiently accurate description of the collision
terms, while the full Fermi--Dirac and Bose--Einstein statistics are
retained where they are relevant for determining $\chi_i(T)$.
This dual approach retains the relevant quantum-statistical effects in the
chemical-potential relations while keeping the collision terms in a
tractable MB form.

At temperatures above approximately $130$~GeV, electroweak sphaleron
processes are active and can redistribute baryon and lepton asymmetries
among the SM particle species, while below this temperature they become
ineffective and are effectively decoupled~\cite{DOnofrio:2014rug}.
A complete treatment would require following the sphaleron-induced
evolution of the relevant chemical potentials across this crossover.
We do not model this dynamics explicitly. Instead, the baryon and lepton
asymmetries entering the chemical-potential relations are treated as
phenomenological inputs, and the sensitivity to the assumed lepton
asymmetry is examined separately in Sec.~\ref{sec:numerical_analysis}.

The baryon asymmetry is well constrained by cosmological observations,
whereas the total lepton asymmetry is comparatively weakly constrained
by cosmological data~\cite{Simha:2008mt,DiClemente:2025awt}.
We therefore consider two representative choices of the lepton asymmetry.
The purpose is not to model the detailed sphaleron dynamics, but to test
whether the main conclusions of the asymmetry-transfer mechanism depend
sensitively on the assumed lepton asymmetry.

We therefore impose the baryon and lepton asymmetries as inputs and
follow their effect on the chemical potentials throughout the temperature
range relevant for the Boltzmann evolution. Assuming electric charge
neutrality, the relevant conservation laws are
\begin{align}
l_f
&\equiv
\frac{\Delta n_f+\Delta n_{\nu_f}}{s},
\label{eq:lepton_asym}
\\
b
&\equiv
\sum_i B_i\frac{\Delta n_i}{s},
\label{eq:baryon_asym}
\\
0
&=
\sum_i Q_i\Delta n_i,
\label{eq:charge_neutrality}
\end{align}
where $s=s(T)$ is the entropy density,
$f\in\{e,\mu,\tau\}$ labels the lepton generations,
$B_i$ denotes the baryon number of species $i$ ($+1/3$ for quarks and
$\tilde{q}$, $-1/3$ for antiquarks and $\tilde{q}^*$, and $0$ otherwise),
and $Q_i$ denotes its electric charge in units of $e$.

Rapid SM interactions with rates satisfying $\Gamma\gg H$ enforce
chemical-equilibrium relations among the corresponding chemical potentials.
In particular, gauge and weak interactions imply
\begin{equation}
\begin{aligned}
\mu_\mu
&=
\mu_e+\mu_{\nu_\mu}-\mu_{\nu_e},
&
\qquad&
(e^-+\nu_\mu\leftrightarrow\mu^-+\nu_e),
\\
\mu_\tau
&=
\mu_e+\mu_{\nu_\tau}-\mu_{\nu_e},
&
\qquad&
(e^-+\nu_\tau\leftrightarrow\tau^-+\nu_e),
\\
\mu_d
&=
\mu_u+\mu_e-\mu_{\nu_e},
&
\qquad&
(u+e^-\leftrightarrow d+\nu_e),
\\
\mu_u
&=
\mu_c=\mu_t,
&
\qquad&
(\text{CKM-mixed weak interactions}),
\\
\mu_d
&=
\mu_s=\mu_b,
&
\qquad&
(\text{CKM-mixed weak interactions}),
\\
\mu_W
&=
\mu_{\nu_e}-\mu_e,
&
\qquad&
(W^+\leftrightarrow e^++\nu_e).
\end{aligned}
\label{eq:chemical_equilibrium_SM}
\end{equation}

Crucially, the chemical potential of $\tilde q$ should not in general be
identified with that of the corresponding SM quark.
Because the portal coupling $\lambda$ is sufficiently small in the
scenario considered here, the conversion processes involving $\tilde q$
need not be fast compared with the Hubble expansion.
We therefore do not impose
$\mu_{\tilde q}=\mu_q$.
Instead, the Boltzmann equations directly determine the evolution of
$\Delta n_{\tilde q}(T)$, while the SM chemical potentials are obtained
from the conservation laws and the SM chemical-equilibrium conditions.
Thus, no independent $\mu_{\tilde q}$ is introduced into the chemical
equilibrium system.

For definiteness, we consider the case in which $\tilde q$ couples to a
quark flavor $q$, and adopt flavor-universal lepton asymmetries,
\begin{equation}
l_e=l_\mu=l_\tau=\frac{l}{3}.
\end{equation}
Substituting the linear relation
$\Delta n_i\simeq\chi_i(T)(\mu_i/T)$ from
Eq.~(\ref{eq:net_density_linear}) into the conservation laws gives the
following system:
\begin{align}
l_e s
&=
\Delta n_e(T,\mu_e)
+
\Delta n_{\nu_e}(T,\mu_{\nu_e}),
\label{eq:cons_le}
\\
l_\mu s
&=
\Delta n_\mu
(T,\mu_e,\mu_{\nu_e},\mu_{\nu_\mu})
+
\Delta n_{\nu_\mu}(T,\mu_{\nu_\mu}),
\label{eq:cons_lmu}
\\
l_\tau s
&=
\Delta n_\tau
(T,\mu_e,\mu_{\nu_e},\mu_{\nu_\tau})
+
\Delta n_{\nu_\tau}(T,\mu_{\nu_\tau}),
\label{eq:cons_ltau}
\\
3bs
&=
\Delta n_u(T,\mu_u)
+
\Delta n_d
(T,\mu_u,\mu_e,\mu_{\nu_e})
+
\Delta n_c(T,\mu_u)
\notag\\
&\quad
+
\Delta n_s
(T,\mu_u,\mu_e,\mu_{\nu_e})
+
\Delta n_t(T,\mu_u)
+
\Delta n_b
(T,\mu_u,\mu_e,\mu_{\nu_e})
+
\Delta n_{\tilde q}(T),
\label{eq:cons_baryon}
\\
0
&=
\frac{2}{3}
\left[
\Delta n_u(T,\mu_u)
+
\Delta n_c(T,\mu_u)
+
\Delta n_t(T,\mu_u)
\right]
\notag\\
&\quad
-
\frac{1}{3}
\left[
\Delta n_d
(T,\mu_u,\mu_e,\mu_{\nu_e})
+
\Delta n_s
(T,\mu_u,\mu_e,\mu_{\nu_e})
+
\Delta n_b
(T,\mu_u,\mu_e,\mu_{\nu_e})
\right]
\notag\\
&\quad
-
\Delta n_e(T,\mu_e)
-
\Delta n_\mu
(T,\mu_e,\mu_{\nu_e},\mu_{\nu_\mu})
-
\Delta n_\tau
(T,\mu_e,\mu_{\nu_e},\mu_{\nu_\tau})
\notag\\
&\quad
+
\Delta n_W(T,\mu_e,\mu_{\nu_e})
+
Q_{\tilde q}\Delta n_{\tilde q}(T).
\label{eq:cons_charge}
\end{align}
Here,
\begin{equation}
\Delta n_W
\equiv
n_{W^+}-n_{W^-},
\end{equation}
and $Q_{\tilde q}$ denotes the electric charge of $\tilde q$ in units of
$e$.
Specifically,
\begin{equation}
Q_{\tilde q}
=
\begin{cases}
+2/3, & \tilde q\ \text{couples to an up-type quark},\\[2mm]
-1/3, & \tilde q\ \text{couples to a down-type quark}.
\end{cases}
\end{equation}

The above equations form a closed linear system for the five independent
SM chemical potentials,
\begin{equation}
\left\{
\mu_u,\,
\mu_e,\,
\mu_{\nu_e},\,
\mu_{\nu_\mu},\,
\mu_{\nu_\tau}
\right\},
\end{equation}
at each temperature $T$.
Solving this system gives the chemical potential relevant for the
conversion processes in Sec.~\ref{subsec:boltzmann_equations} in the form
\begin{equation}
\frac{\mu_q}{T}
=
A_q(T)\,b
+
B_q(T)\,l
+
C_q(T)\,Y_\Delta,
\label{eq:muq_solution}
\end{equation}
where
\begin{equation}
Y_\Delta
=
\frac{\Delta n_{\tilde q}(T)}{s}
=
Y_{\tilde q}-Y_{\tilde q^*}.
\end{equation}
Here $\mu_q$ denotes the chemical potential of the quark flavor associated
with $\tilde q$.
In particular,
$\mu_q=\mu_u$ when $\tilde q$ couples to an up-type quark, while
$\mu_q=\mu_d$ when it couples to a down-type quark.
The coefficients $A_q(T)$, $B_q(T)$, and $C_q(T)$ are dimensionless,
temperature-dependent functions determined algebraically by the conservation
laws above.
They incorporate the temperature dependence of the coefficients
$\chi_i(T)$ and hence the effects associated with the relevant
mass-to-temperature ratios.

Equation~(\ref{eq:muq_solution}) provides the required closure of the
Boltzmann system.
In particular, the chemical-potential factors appearing in the effective
equilibrium abundances introduced in
Sec.~\ref{subsec:boltzmann_equations},
\begin{equation}
Y_{\tilde q,\mathrm{w}}^{\mathrm{eq}}
=
Y_{\tilde q,\mathrm{w/o}}^{\mathrm{eq}}
e^{+\mu_q/T},
\qquad
Y_{\tilde q^*,\mathrm{w}}^{\mathrm{eq}}
=
Y_{\tilde q^*,\mathrm{w/o}}^{\mathrm{eq}}
e^{-\mu_q/T},
\end{equation}
can now be evaluated directly from $b$, $l$, $Y_\Delta$, and the
temperature $T$.

For our baseline numerical analysis, we take the baryon asymmetry to be
\begin{equation}
b=9\times10^{-11}.
\end{equation}
As a benchmark with a lepton asymmetry comparable to the baryon asymmetry,
we take
\begin{equation}
l=-b.
\end{equation}
To explore the sensitivity to a substantially larger lepton asymmetry, we
also consider
\begin{equation}
l_e=l_\mu=l_\tau=10^{-4},
\qquad
l=3\times10^{-4},
\label{eq:larger_lepton_asymmetry}
\end{equation}
following Ref.~\cite{Schwarz:2009ii}.
The numerical impact of these two choices on the asymmetry-transfer
mechanism is discussed in Sec.~\ref{sec:numerical_analysis}.

\section{Numerical Results}
\label{sec:numerical_analysis}

We now turn to the numerical evolution of the coupled Boltzmann equations
derived in Sec.~\ref{subsec:boltzmann_equations}.  Our main goal is to
demonstrate explicitly the asymmetry-transfer mechanism illustrated
schematically in Fig.~\ref{fig:conceptual_plot} and to identify the region of
parameter space in which the pre-existing asymmetry stored in the partner
sector provides a significant contribution to the dark matter relic
abundance.

Throughout this section, we take
\begin{equation}
b=9\times10^{-11},
\end{equation}
and use the two representative choices of the lepton asymmetry introduced
in Sec.~\ref{sec:chemical_potential},
\begin{equation}
l=-b
\end{equation}
and
\begin{equation}
l=3\times10^{-4}.
\end{equation}
The comparison between these two choices provides a robustness test of the
asymmetry-transfer mechanism with respect to the assumed lepton asymmetry.

For the benchmark evolution plots, we consider a stop-like partner
$\tilde q=\tilde t$ with
\begin{equation}
m_{\tilde q}=2~\mathrm{TeV},
\qquad
m_\chi=500~\mathrm{GeV},
\end{equation}
and use the $s$-wave-dominated $\chi\chi$ effective annihilation cross
section
\begin{equation}
C_s=6\times10^{-9}~\mathrm{GeV}^{-2},
\qquad
C_p=0.
\end{equation}
In the absence of asymmetry transfer, these parameters give
\begin{equation}
\Omega_\chi h^2\simeq0.04,
\end{equation}
so that the ordinary thermal relic of $\chi$ provides only a subdominant
contribution to the observed dark matter abundance.

\subsection{Benchmark evolution and the asymmetry-transfer mechanism}
\label{subsec:benchmark_evolution}

\begin{figure}[t]
\centering
\includegraphics[width=0.48\textwidth]{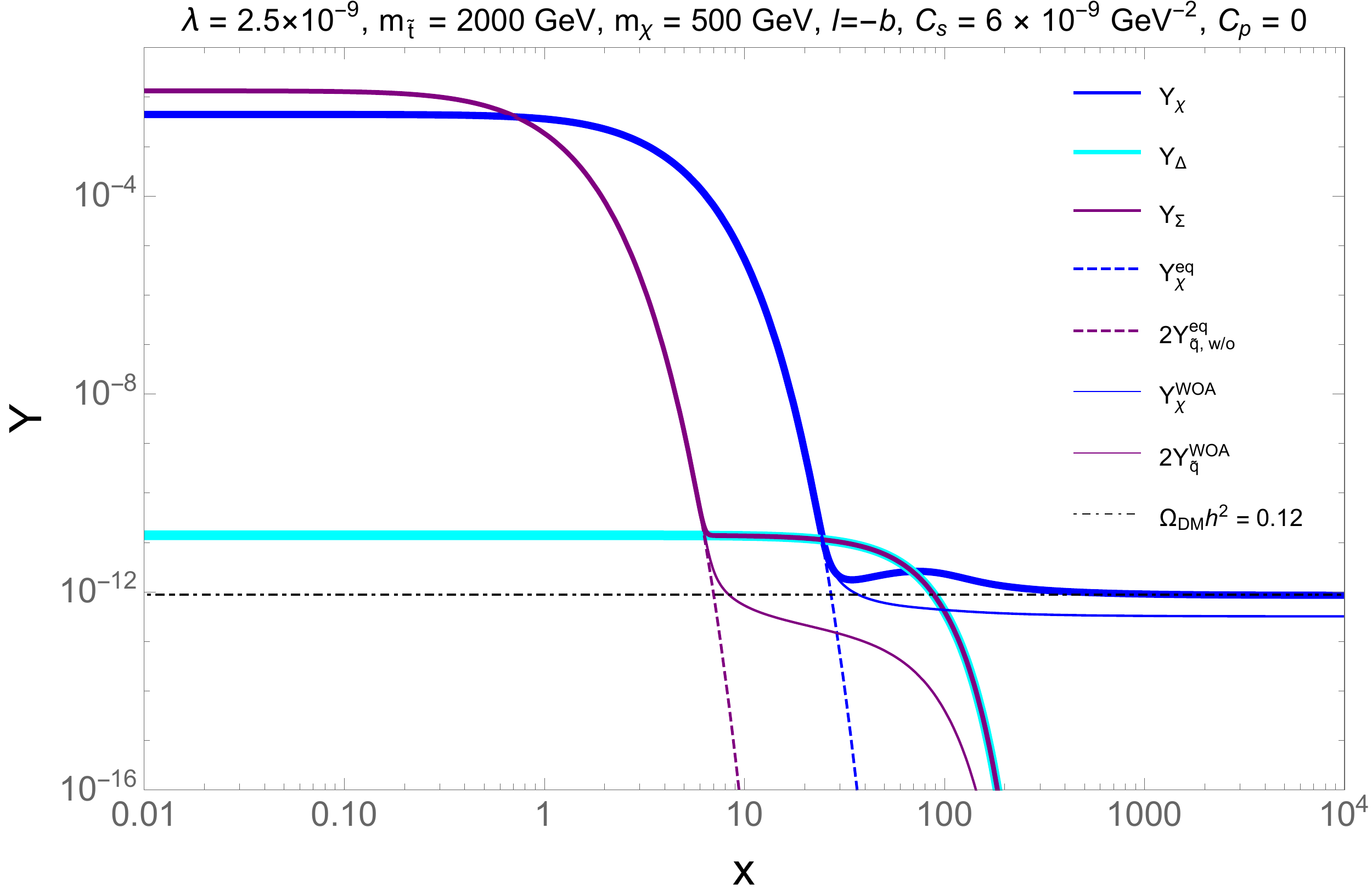}
\hfill
\includegraphics[width=0.48\textwidth]{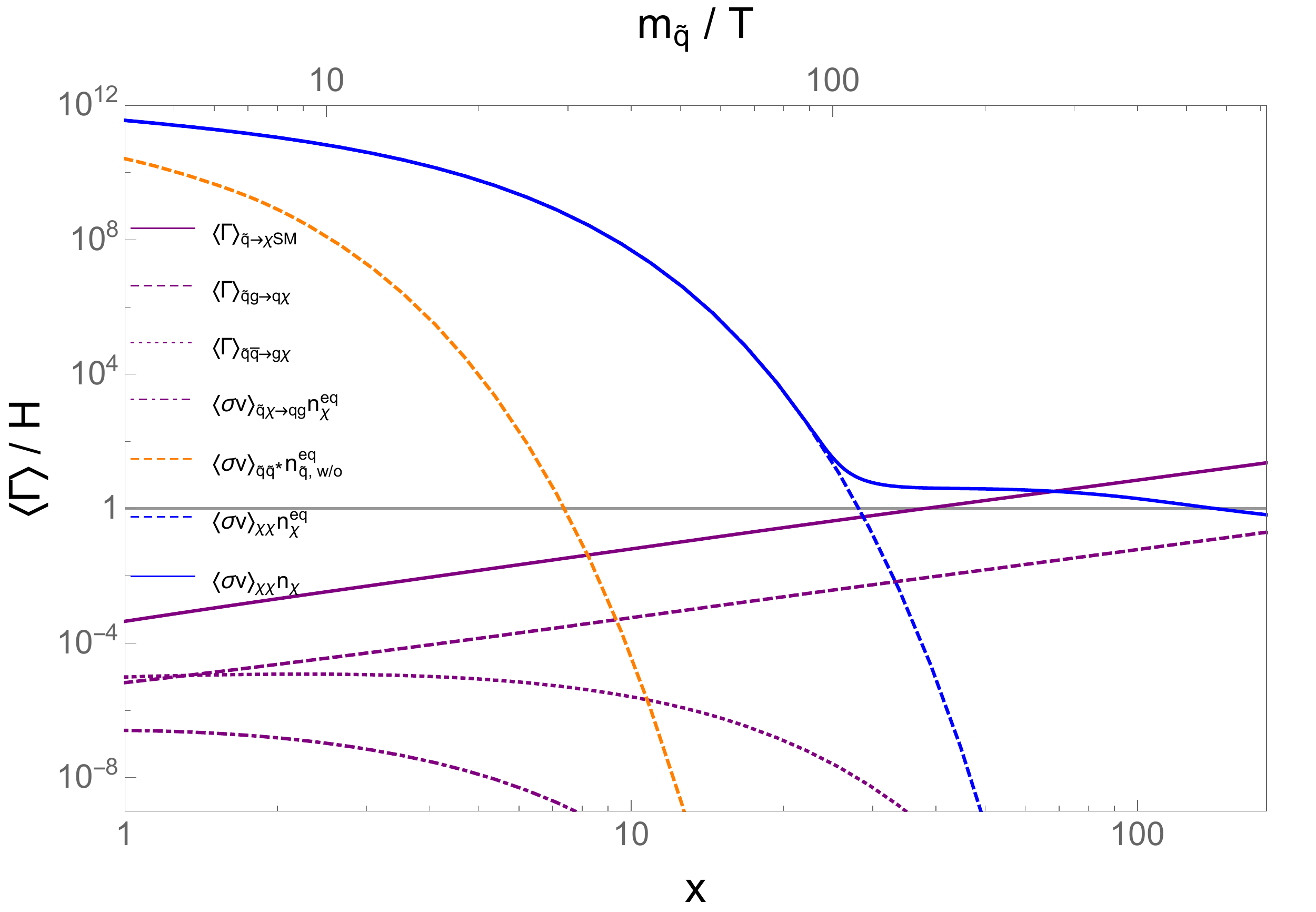}
\caption{
Evolution of the particle abundances and interaction rates for the
benchmark choice
$m_{\tilde q}=2~\mathrm{TeV}$,
$m_\chi=500~\mathrm{GeV}$,
$\lambda=2.5\times10^{-9}$,
$l=-b$,
$C_s=6\times10^{-9}~\mathrm{GeV}^{-2}$,
and $C_p=0$.
Left panel: the yields $Y_\chi$, $Y_\Sigma$, and $Y_\Delta$ obtained from
the coupled Boltzmann equations, together with the corresponding
equilibrium and ``without asymmetry'' results (labeled with ``WOA'').
The WOA results are obtained by setting $\mu_q=0$ and $Y_\Delta=0$ when
solving Eqs.~(\ref{eq:BE_chi}) and (\ref{eq:BE_Sigma}).
The horizontal dot-dashed line indicates the observed dark matter abundance
$\Omega_{\rm DM}h^2=0.12$.
Right panel: the relevant thermally averaged interaction rates divided by
the Hubble expansion rate as functions of $x=m_\chi/T$.
The upper axis indicates the corresponding value of $m_{\tilde q}/T$.
}
\label{fig:benchmark_evolution}
\end{figure}

We first consider
\begin{equation}
\lambda=2.5\times10^{-9},
\end{equation}
for which the final dark matter abundance is close to the observed value.
The evolution of the relevant yields and interaction rates is shown in
Fig.~\ref{fig:benchmark_evolution}.

The benchmark evolution illustrates the three main stages underlying the
asymmetry-transfer mechanism in Fig.~\ref{fig:conceptual_plot}.  At early times, both $\chi$
and $\tilde q$ are in thermal equilibrium, while the primordial asymmetry
is shared between the SM quark and partner sectors.  The subsequent
$\tilde q\tilde q^*$ QCD annihilation efficiently removes the
particle-antiparticle symmetric component of the partner population.
Since these annihilations involve one $\tilde q$ and one $\tilde q^*$,
they do not directly change the net asymmetry
\begin{equation}
Y_\Delta=Y_{\tilde q}-Y_{\tilde q^*}.
\end{equation}
Consequently, after the symmetric component has been depleted, the
remaining partner abundance becomes increasingly dominated by the
asymmetric component.

For the benchmark masses, the depletion of the symmetric partner
population occurs around
\begin{equation}
\frac{m_{\tilde q}}{T}\sim25,
\qquad
x\sim6.
\end{equation}
At early times, the conversion processes are not sufficiently effective to
maintain chemical equilibrium between $\tilde q$ and $\chi$.
As the Universe evolves, the ratio of the relevant conversion rates to the
Hubble expansion rate changes, and the conversion processes eventually
become dynamically important.
For the benchmark point, this occurs after
the ordinary thermal freeze-out of $\chi$ has taken place.  The subsequent
conversion therefore transfers part of the asymmetry stored in the
$\tilde q$ sector to $\chi$, with the decay
$\tilde q\to\chi q$ providing the dominant contribution.

The conversion processes therefore do not simply maintain chemical
equilibrium between $\tilde q$ and $\chi$ throughout the evolution.
Instead, their time-dependent efficiency determines how much of the
asymmetry stored in the partner reservoir is transferred to $\chi$ and,
more importantly, when this transfer takes place relative to the freeze-out
of $\chi$.

As the asymmetry is transferred, $Y_\chi$ temporarily increases above the
would-be thermal relic trajectory.  The newly produced $\chi$ particles
can still annihilate efficiently for part of this period, leading to a
subsequent decrease of $Y_\chi$.  The final relic abundance is therefore
determined by the competition between the late-time injection of $\chi$
from the surviving partner asymmetry and the residual $\chi\chi$
annihilation rate.

This behavior provides a direct numerical realization of the conceptual
picture in Fig.~\ref{fig:conceptual_plot}: QCD annihilations first remove
the symmetric component of the partner population while preserving its
net asymmetry.  The surviving asymmetric component is subsequently
transferred to $\chi$ at relatively late times, after the ordinary thermal
freeze-out of $\chi$.  The final dark matter abundance is consequently
controlled by both the amount of asymmetry that survives in the partner
sector and the efficiency with which the injected $\chi$ particles
annihilate after the transfer.

\subsection{Dependence on the portal coupling}
\label{subsec:lambda_dependence}

The portal coupling controls the time at which the partner asymmetry is
transferred to $\chi$.  Since the conversion rates scale approximately as
$\lambda^2$, varying $\lambda$ changes the relative timing of asymmetry
transfer and the ordinary thermal freeze-out of $\chi$.

\begin{figure}[t]
\centering
\includegraphics[width=0.48\textwidth]{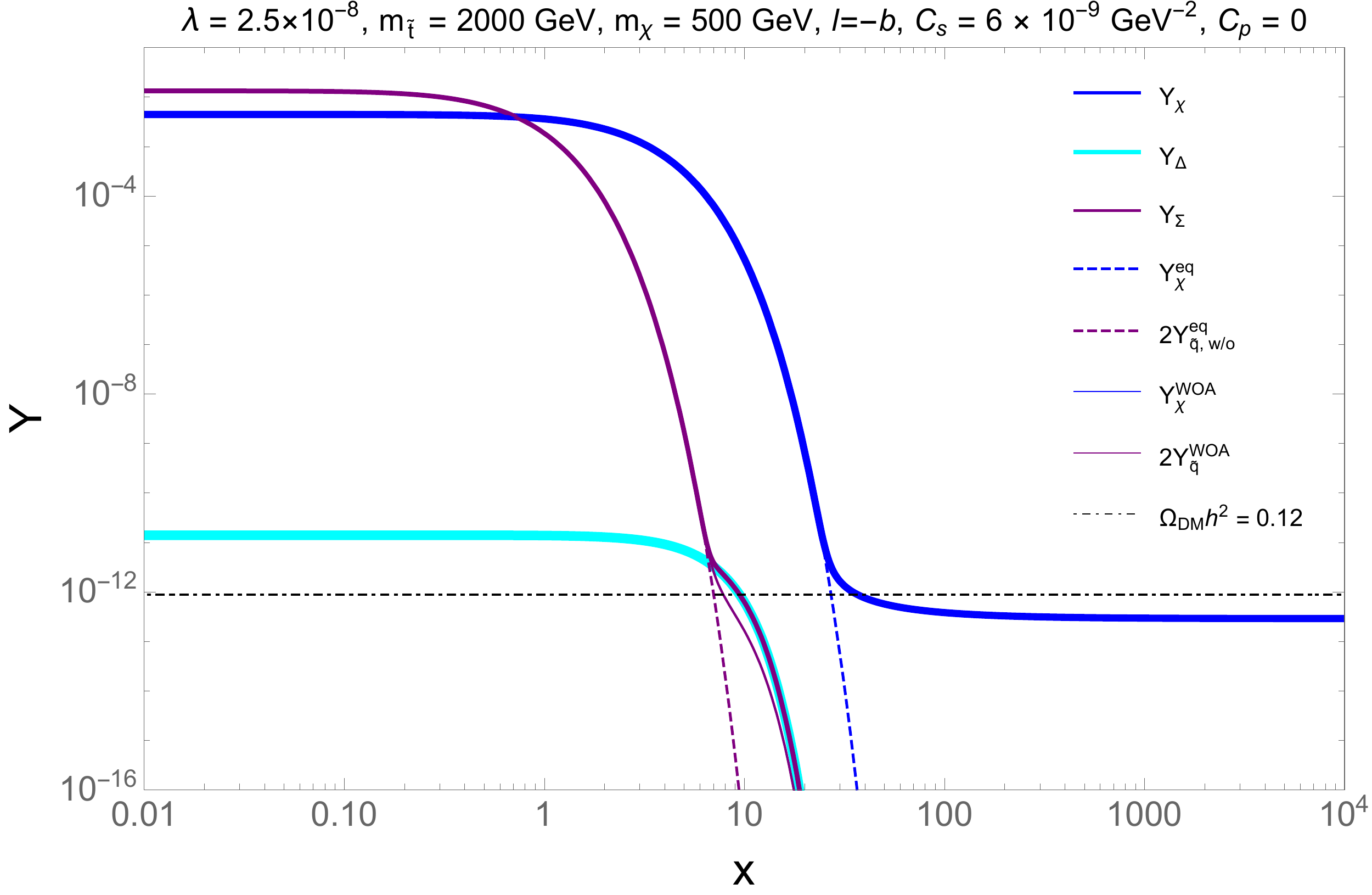}
\hfill
\includegraphics[width=0.48\textwidth]{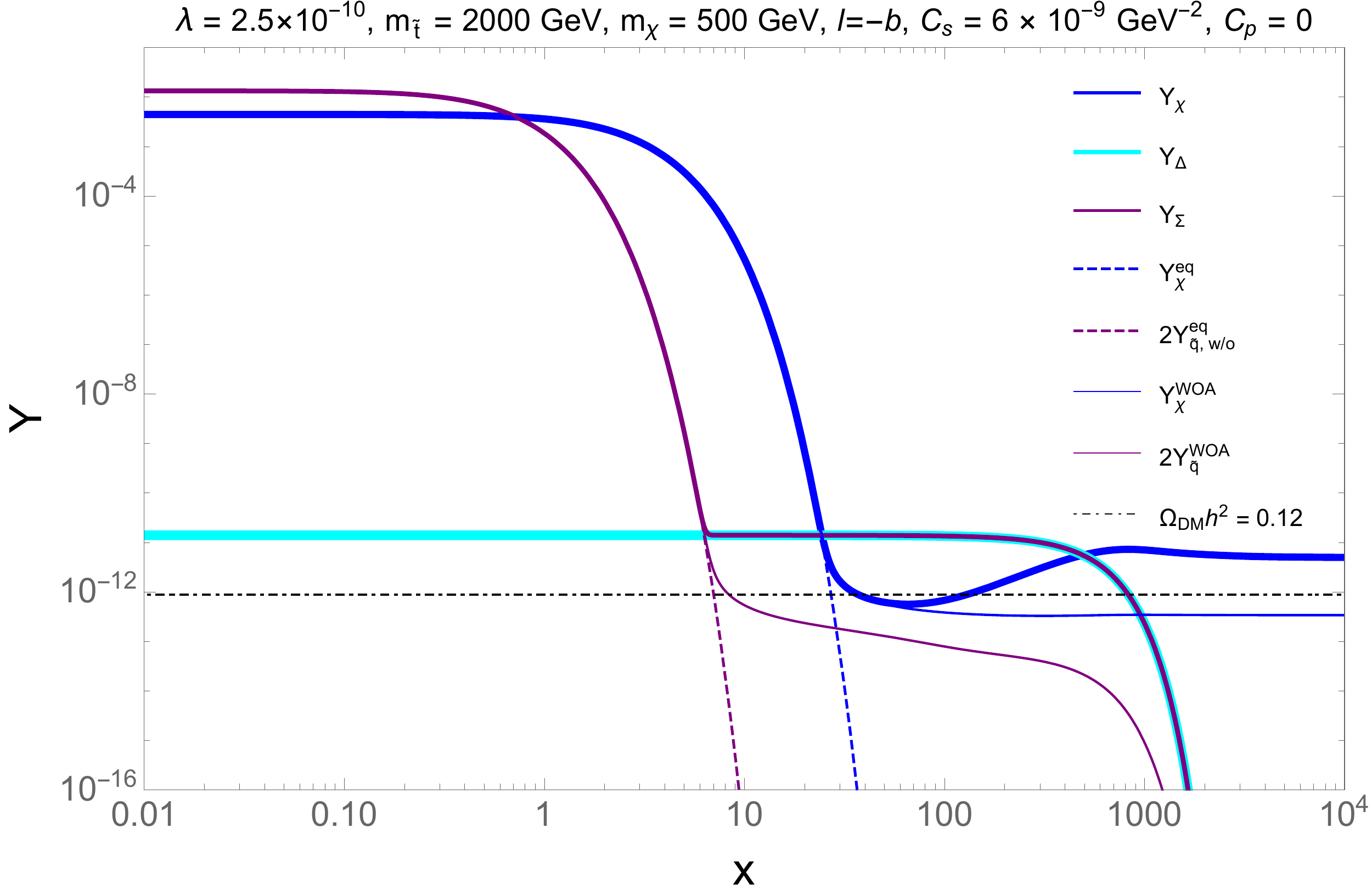}
\caption{
Evolution of the particle abundances for the same benchmark parameters as
Fig.~\ref{fig:benchmark_evolution}, but with
$\lambda=2.5\times10^{-8}$ (left) and
$\lambda=2.5\times10^{-10}$ (right).
The larger coupling leads to an earlier transfer of the partner-sector
asymmetry, whereas the smaller coupling delays the transfer until the
residual $\chi\chi$ annihilation becomes less efficient.
}
\label{fig:lambda_evolution}
\end{figure}

For a larger coupling,
\begin{equation}
\lambda=2.5\times10^{-8},
\end{equation}
the conversion processes become effective at an earlier stage of the
evolution.  As a result, the partner-sector asymmetry is transferred to
$\chi$ before the ordinary thermal freeze-out of $\chi$ is completed.
The $\chi$ particles produced through this early transfer can still
efficiently annihilate, so that the transferred component does not provide
a significant additional contribution to the final dark matter abundance.
The relic density therefore remains close to that obtained from the
ordinary thermal freeze-out of $\chi$ alone.

In the opposite direction, for
\begin{equation}
\lambda=2.5\times10^{-10},
\end{equation}
the weaker portal interaction delays the transfer of the partner-sector
asymmetry.  A larger fraction of the surviving asymmetry is consequently
transferred to $\chi$ only at later times, when the $\chi\chi$ annihilation
rate has already become substantially smaller than the Hubble rate.  The
newly produced $\chi$ particles are therefore less efficiently
annihilated, leading to a larger final relic abundance.

These three cases illustrate the characteristic dependence of the final
relic abundance on the timing of asymmetry transfer:
\begin{equation}
\lambda\ {\rm too\ large}
\quad\Longrightarrow\quad
{\rm transfer\ too\ early},
\end{equation}
\begin{equation}
\lambda\ {\rm intermediate}
\quad\Longrightarrow\quad
{\rm transfer\ during\ the\ relevant\ freeze\!-\!out\ epoch},
\end{equation}
and
\begin{equation}
\lambda\ {\rm too\ small}
\quad\Longrightarrow\quad
{\rm transfer\ too\ late}.
\end{equation}
For early transfer, the injected $\chi$ particles are efficiently
annihilated, whereas for sufficiently late transfer the annihilation
efficiency is too small to remove the injected population.  An
intermediate portal coupling therefore allows the asymmetry to be
transferred at a time such that its contribution to the final relic
abundance can be significant and can reproduce the observed value.

\begin{figure}[t]
\centering
\includegraphics[width=0.48\textwidth]{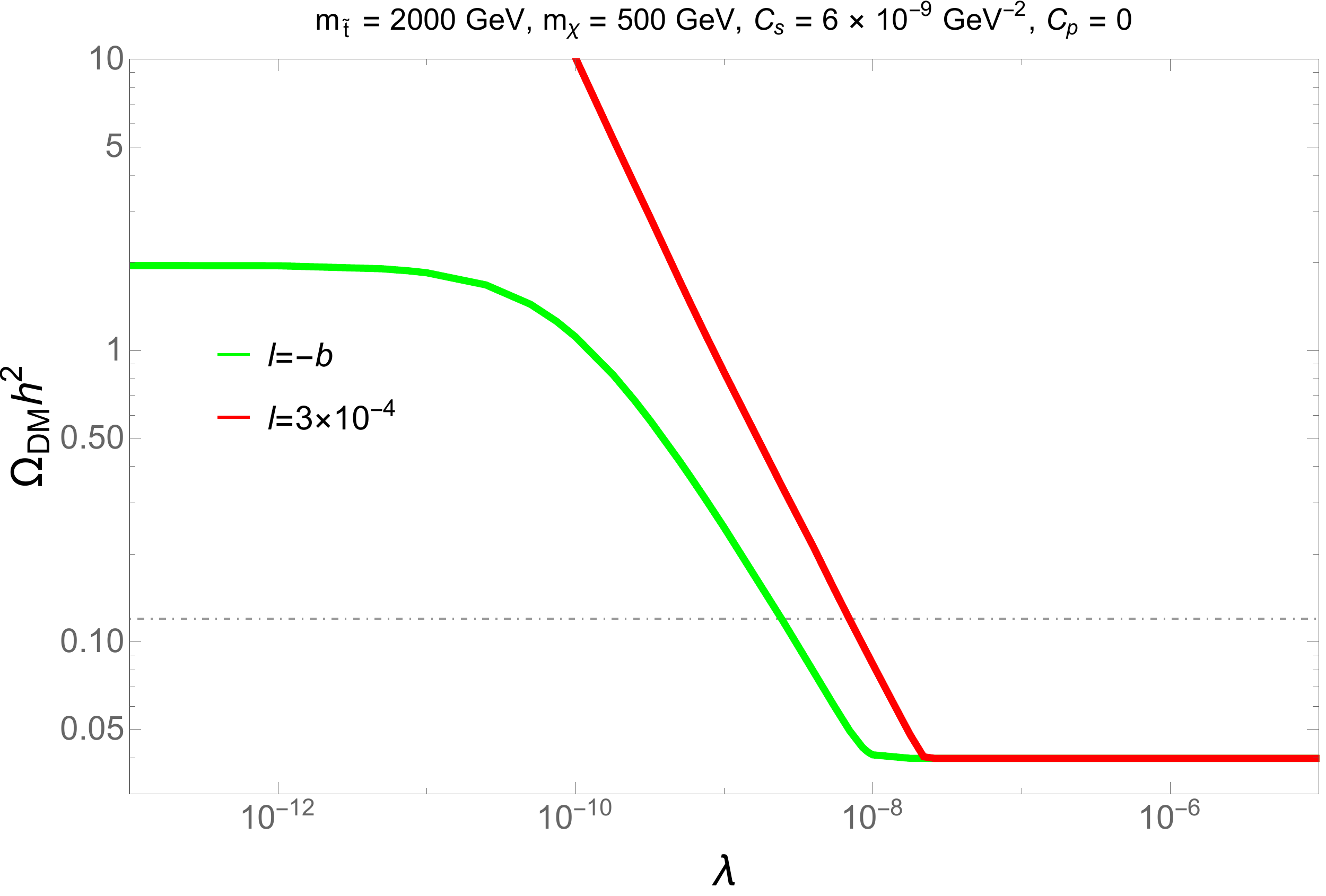}
\hfill
\includegraphics[width=0.48\textwidth]{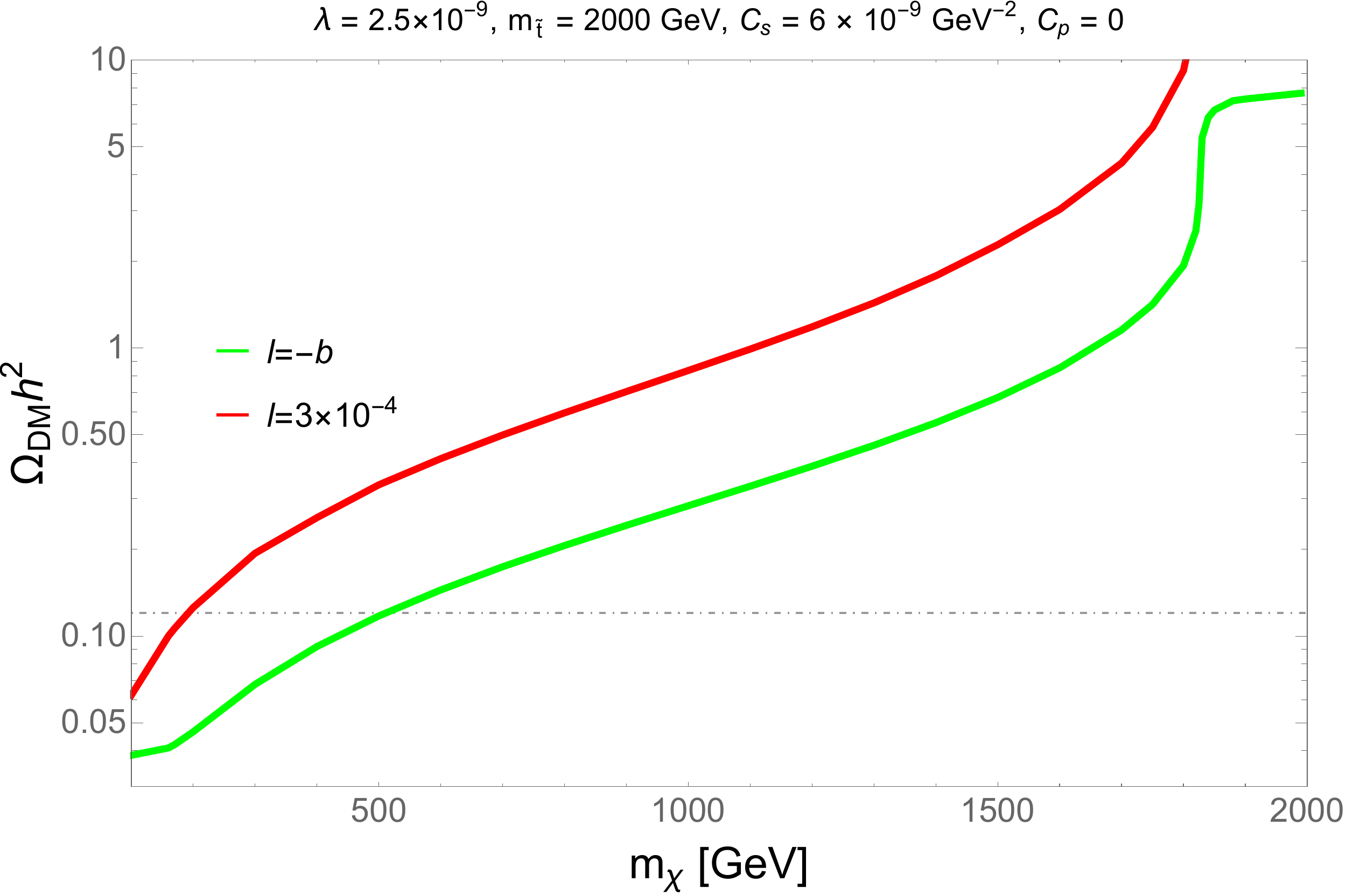}
\caption{
Final dark matter relic abundance $\Omega_{\rm DM}h^2$ as a function of
the portal coupling $\lambda$ (left) and of the dark matter mass
$m_\chi$ (right), for fixed
$m_{\tilde q}=2~\mathrm{TeV}$.
The green and red curves correspond to
$l=-b$ and $l=3\times10^{-4}$, respectively.
The remaining parameters are those of the benchmark in
Fig.~\ref{fig:benchmark_evolution}.
The horizontal line indicates the observed value
$\Omega_{\rm DM}h^2\simeq0.12$.
}
\label{fig:relic_lambda_mchi}
\end{figure}

For the benchmark mass choice, a scan over $\lambda$ confirms this
qualitative picture, as shown in the left panel of
Fig.~\ref{fig:relic_lambda_mchi}. 
For $\lambda\gtrsim10^{-8}$, the asymmetry-transfer contribution becomes
negligible, while for sufficiently small $\lambda$ the transferred
asymmetry can lead to an overabundance.  

At even smaller couplings, the late transfer is accompanied by an
increasingly long lifetime of the partner, and cosmological constraints on
long-lived colored particles become relevant.
In particular, requiring the partner to decay before Big Bang
nucleosynthesis~\cite{Cyburt:2009pg,Kawasaki:2020qxm,Bianco:2026dvc} gives a characteristic lower bound of order
$\lambda\gtrsim10^{-13}$ for the benchmark considered here.

The right panel of Fig.~\ref{fig:relic_lambda_mchi} shows the dependence of
the final relic abundance on $m_\chi$, with
$\lambda=2.5\times10^{-9}$ and $m_{\tilde q}=2~\mathrm{TeV}$ fixed.
The relic abundance generally increases as $m_\chi$ is increased.
The two representative choices of the lepton asymmetry give similar
overall behavior, although their numerical values differ because of the
different quark chemical potentials entering the conversion terms.

For the stop-like benchmark, an additional feature appears when
\begin{equation}
m_{\tilde t}-m_\chi < m_t,
\end{equation}
where the two-body decay
$\tilde t\to\chi t$ becomes kinematically forbidden.  The relic-density
curve changes more rapidly across this threshold because the dominant
conversion channel changes, with scattering and higher-body decay
processes becoming increasingly important.  As $m_\chi$ approaches
$m_{\tilde t}$, the available phase space for the conversion processes is
further reduced, and the relic abundance approaches a limiting behavior.

\subsection{Dependence on the mass hierarchy}
\label{subsec:mass_dependence}

The dependence on $m_\chi$ shown in the right panel of
Fig.~\ref{fig:relic_lambda_mchi} reflects the interplay between the mass
hierarchy and the timing of asymmetry transfer.  We therefore examine this
dependence more directly by varying $m_\chi$ while keeping
\begin{equation}
m_{\tilde q}=2~\mathrm{TeV},
\qquad
\lambda=2.5\times10^{-9},
\end{equation}
fixed.

\begin{figure}[t]
\centering
\includegraphics[width=0.48\textwidth]{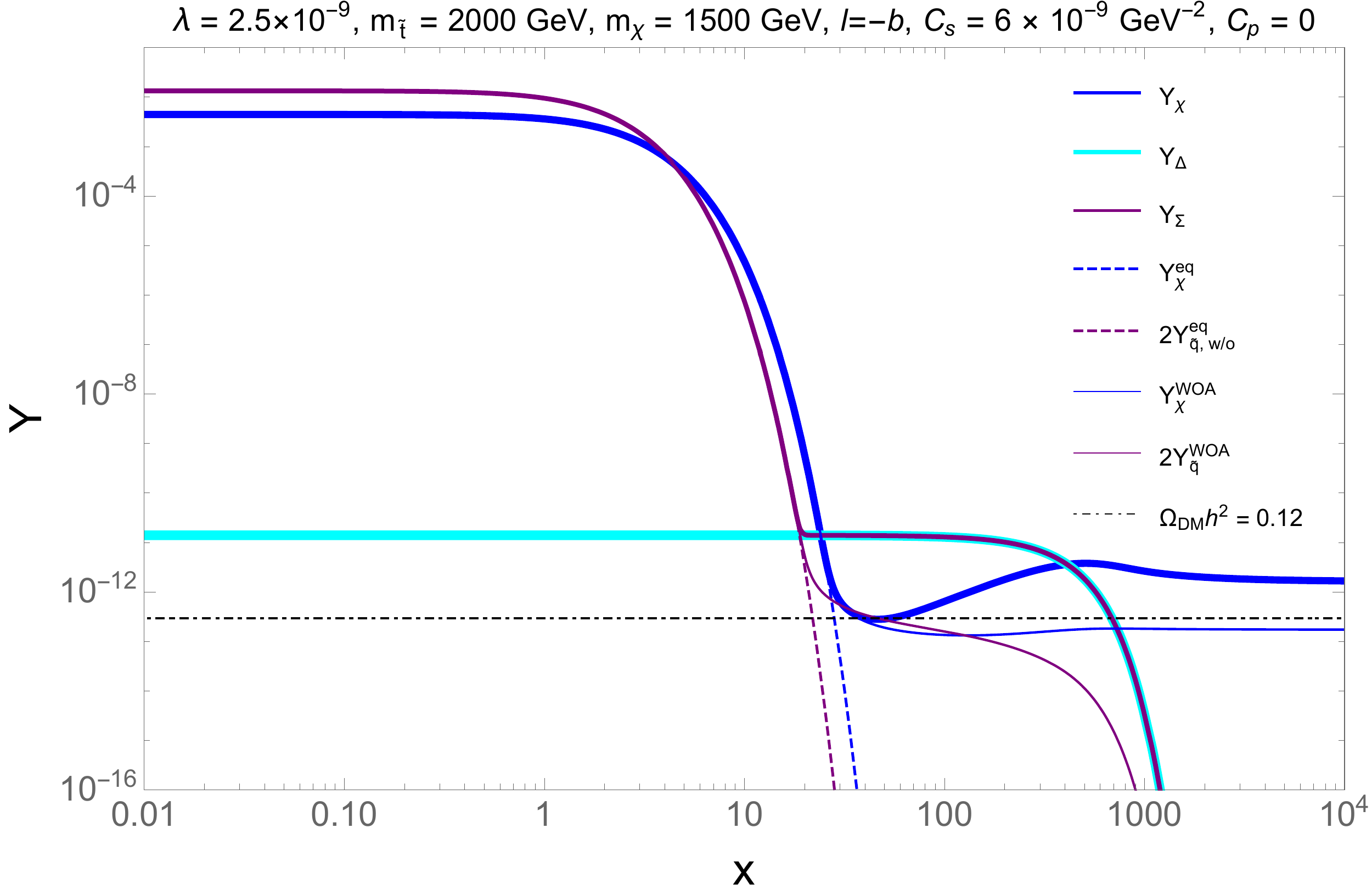}
\hfill
\includegraphics[width=0.48\textwidth]{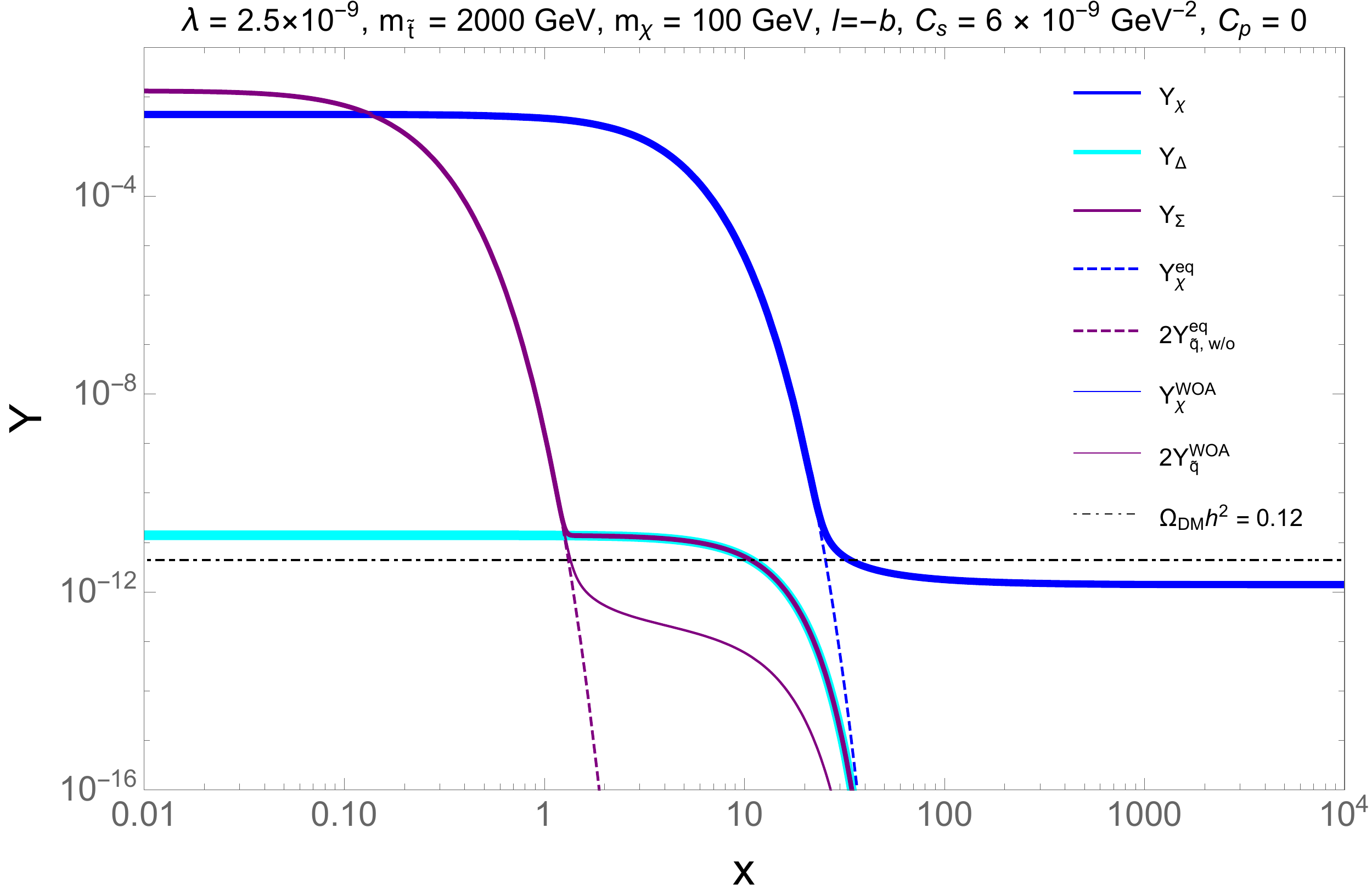}
\caption{
Evolution of the particle abundances for
$m_\chi=1.5~\mathrm{TeV}$ (left) and
$m_\chi=100~\mathrm{GeV}$ (right), with
$m_{\tilde q}=2~\mathrm{TeV}$ and
$\lambda=2.5\times10^{-9}$.
The remaining parameters are the same as in
Fig.~\ref{fig:benchmark_evolution}.
The comparison illustrates how the mass hierarchy controls the relative
timing of asymmetry transfer and $\chi$ freeze-out.
}
\label{fig:mass_evolution}
\end{figure}

When $m_\chi$ is increased toward $m_{\tilde q}$, the relative timing of
the conversion and $\chi$ freeze-out changes, and the surviving partner
asymmetry can be transferred at larger values of
$x=m_\chi/T$.
At these later times, the residual $\chi\chi$ annihilation is less
efficient, enhancing the contribution of the transferred asymmetry to the
final relic density.

For a much smaller $m_\chi$, the hierarchy
\begin{equation}
m_{\tilde q}\gg m_\chi
\end{equation}
separates the QCD depletion of the partner population, the conversion of
the surviving asymmetry, and the ordinary freeze-out of $\chi$ more
strongly.  In this regime, the asymmetry transfer can occur sufficiently
early that much of the resulting $\chi$ population is subsequently
annihilated, reducing its impact on the final relic abundance.

The important point is that the mechanism does not require a compressed
spectrum.  In contrast to ordinary coannihilation, the relevant reservoir
is the surviving asymmetric component rather than the thermally suppressed
symmetric abundance of the heavier state.  A substantial mass hierarchy
can therefore remain compatible with efficient asymmetry transfer, provided
the conversion history is appropriately arranged.

\begin{figure}[t]
\centering
\includegraphics[width=0.72\textwidth]{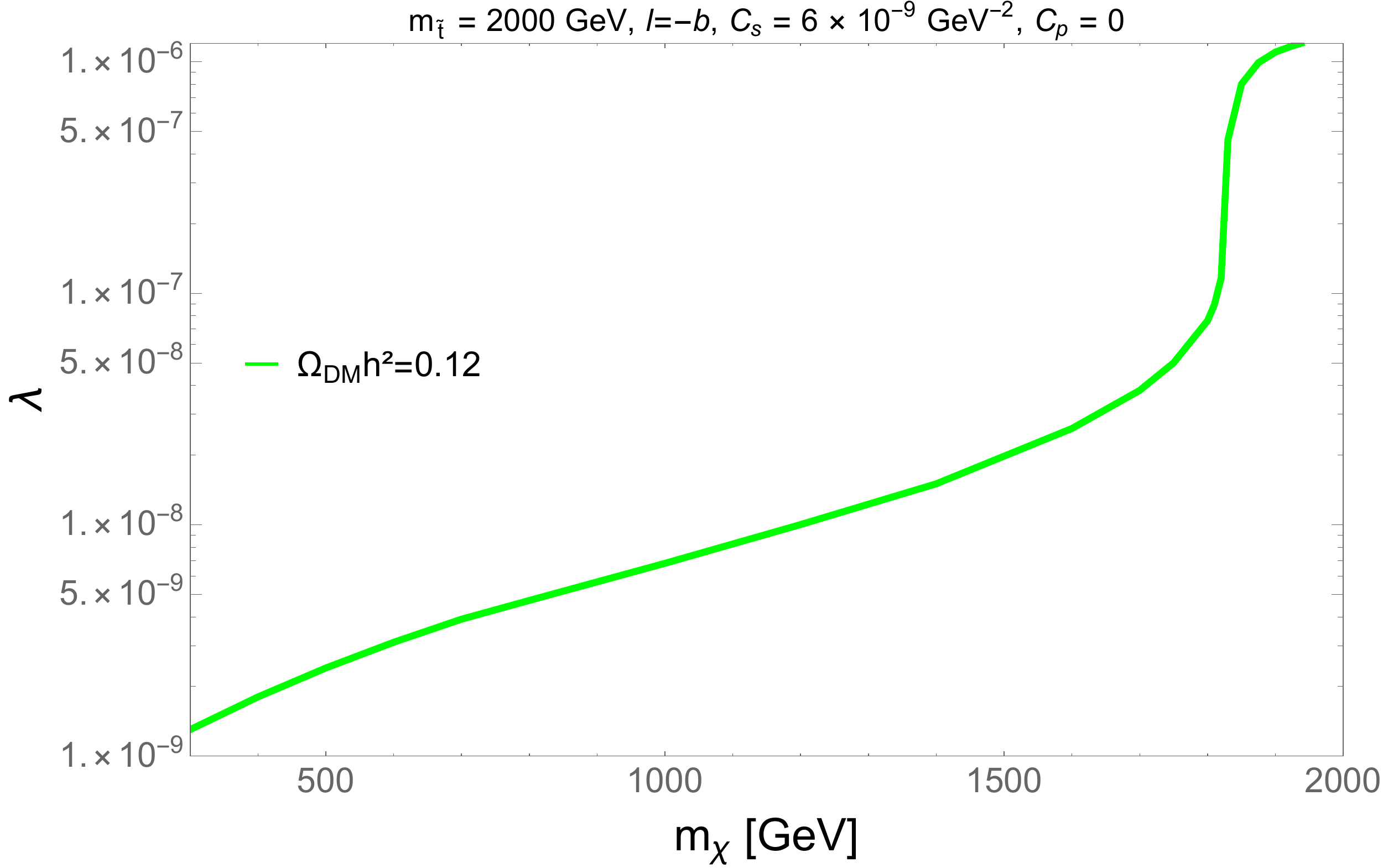}
\caption{
Parameter combinations in the
$(m_\chi,\lambda)$ plane that give
$\Omega_{\rm DM}h^2\simeq0.12$ for fixed
$m_{\tilde q}=2~\mathrm{TeV}$, $l=-b$, and the benchmark
$s$-wave annihilation cross section.
The curve illustrates the degeneracy between the portal coupling and the
dark matter mass in obtaining the observed relic abundance.
}
\label{fig:relic_contour}
\end{figure}

More generally, the relic abundance depends on both the mass hierarchy and
the portal coupling.  Different combinations of
$(\lambda,m_{\tilde q},m_\chi)$ can therefore lead to the same observed
dark matter abundance, as illustrated by the contour in
Fig.~\ref{fig:relic_contour}.

The benchmark results presented above have been obtained mainly for
$m_{\tilde t}=2~\mathrm{TeV}$.  To illustrate the behavior for a different
partner mass, we also consider
\begin{equation}
m_{\tilde t}=3~\mathrm{TeV}.
\end{equation}
For $l=-b$ and
$\langle\sigma v\rangle_{\chi\chi}=6\times10^{-9}~\mathrm{GeV}^{-2}$,
appropriate choices of the portal coupling and dark matter mass can again yield the observed relic abundance.
For example,
\begin{equation}
(\lambda,m_{\tilde t},m_\chi)
=
(1.8\times10^{-9},\,3~\mathrm{TeV},\,480~\mathrm{GeV})
\end{equation}
and
\begin{equation}
(\lambda,m_{\tilde t},m_\chi)
=
(1.0\times10^{-8},\,3~\mathrm{TeV},\,1.6~\mathrm{TeV})
\end{equation}
both give
\begin{equation}
\Omega_\chi h^2\simeq0.12.
\end{equation}
These examples demonstrate that the asymmetry-transfer mechanism can
remain viable for different partner masses, with the portal coupling and
dark matter mass adjusting accordingly.

\subsection{Near-degenerate spectra}
\label{subsec:near_degenerate}

In the hierarchical parameter region relevant for the main asymmetry-transfer mechanism, we typically consider $\lambda\lesssim10^{-8}$. Near-degenerate spectra can remain viable for larger portal couplings, for which additional conversion channels become relevant. 
For a stop-like partner, this region has an additional kinematic feature
because the two-body decay
$
\tilde t\to\chi t
$
becomes kinematically forbidden when $m_{\tilde t}-m_\chi<m_t$. 
The conversion dynamics is then controlled by scattering processes and
higher-body decays rather than by the two-body decay alone.

\begin{figure}[t]
\centering
\includegraphics[width=0.48\textwidth]{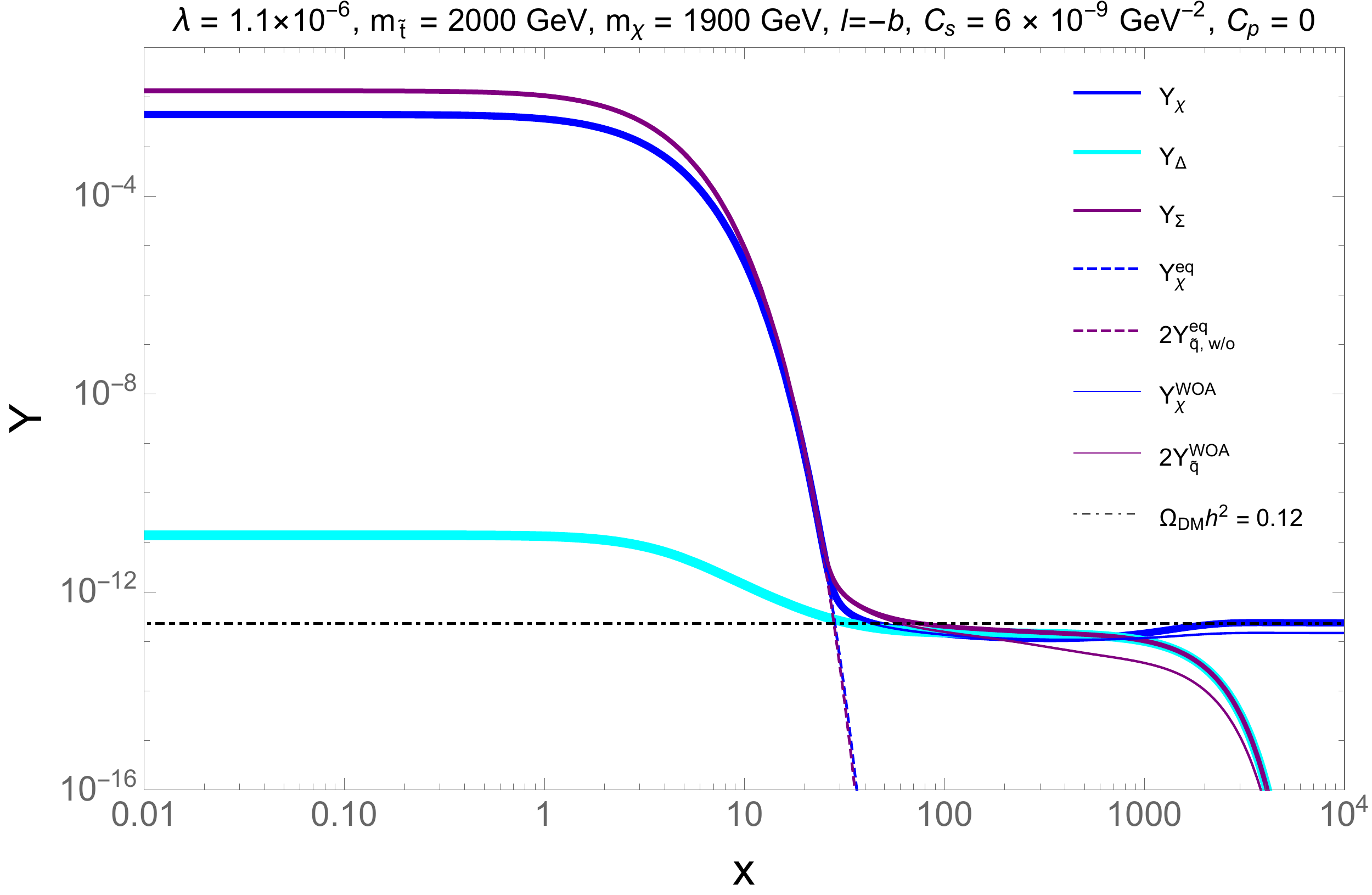}
\hfill
\includegraphics[width=0.48\textwidth]{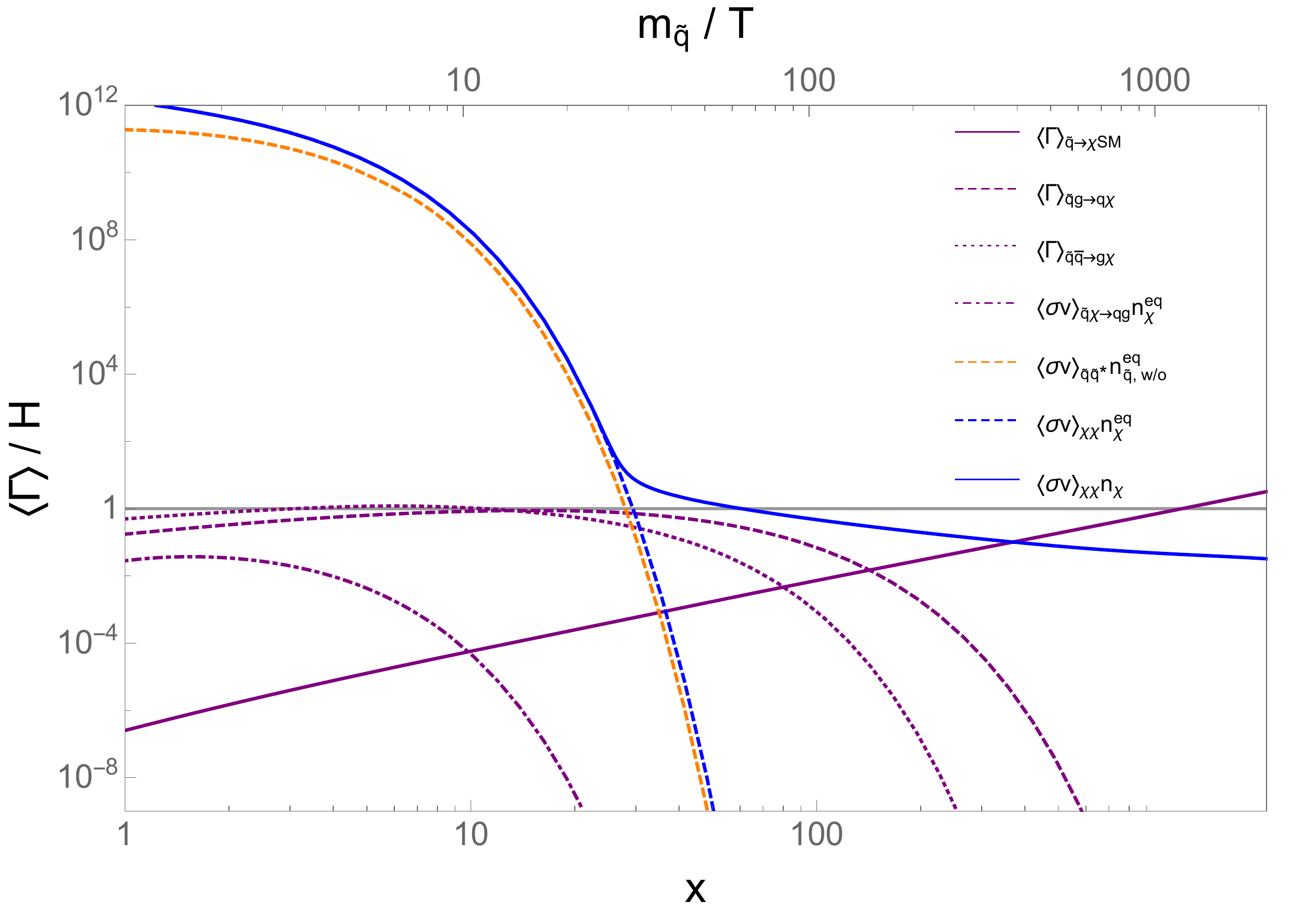}
\caption{
Near-degenerate mass spectrum with
$m_{\tilde t}=2~\mathrm{TeV}$,
$m_\chi=1.9~\mathrm{TeV}$, and
$\lambda=1.1\times10^{-6}$.
Left panel: evolution of the particle abundances.
Right panel: the corresponding conversion and annihilation rates relative
to the Hubble expansion rate.
Because the two-body decay $\tilde t\to\chi t$ is kinematically forbidden,
the conversion proceeds through scattering processes at earlier times and
higher-body decays at later times.
}
\label{fig:near_degenerate}
\end{figure}

In this regime, the conversion history can contain more than one stage.
The scattering processes involving thermal top quarks become relevant at
early times but are rapidly suppressed as the temperature decreases,
owing both to the top-quark mass and to the small available phase space.
The higher-body decay channels can then become important at later times.
The remaining partner asymmetry is consequently transferred in more than
one stage, with part of the asymmetry being processed before $\chi$
freeze-out and the remainder being transferred later.
This near-degenerate regime is therefore qualitatively distinct from the
small-$\lambda$ hierarchical regime discussed above.

This example demonstrates that the asymmetry-transfer mechanism is not
restricted to a single dominant microscopic conversion channel.  The
relevant requirement is instead that the total conversion history leaves
an appropriate fraction of the partner-sector asymmetry to be transferred
at a time when it can contribute to the final $\chi$ abundance.

\subsection{Dependence on the dark matter annihilation mode}
\label{subsec:annihilation_dependence}

The final relic abundance is sensitive to the annihilation properties of
$\chi$ when the asymmetry transfer occurs after, or around, the ordinary
thermal freeze-out of $\chi$.  In this case, the injected $\chi$ particles
are no longer efficiently driven back toward chemical equilibrium, and
their survival in the dark matter sector depends on the $\chi\chi$
annihilation rate at the time of injection.  Consequently, the relic
abundance can depend sensitively on whether the annihilation is $s$-wave or
$p$-wave dominated.

By contrast, if the asymmetry is transferred sufficiently early, before
$\chi$ freeze-out, the injected $\chi$ particles can still efficiently
annihilate and track the thermal abundance.  In this regime, the final
relic density is much less sensitive to the detailed annihilation mode.

\begin{figure}[t]
\centering
\includegraphics[width=0.48\textwidth]{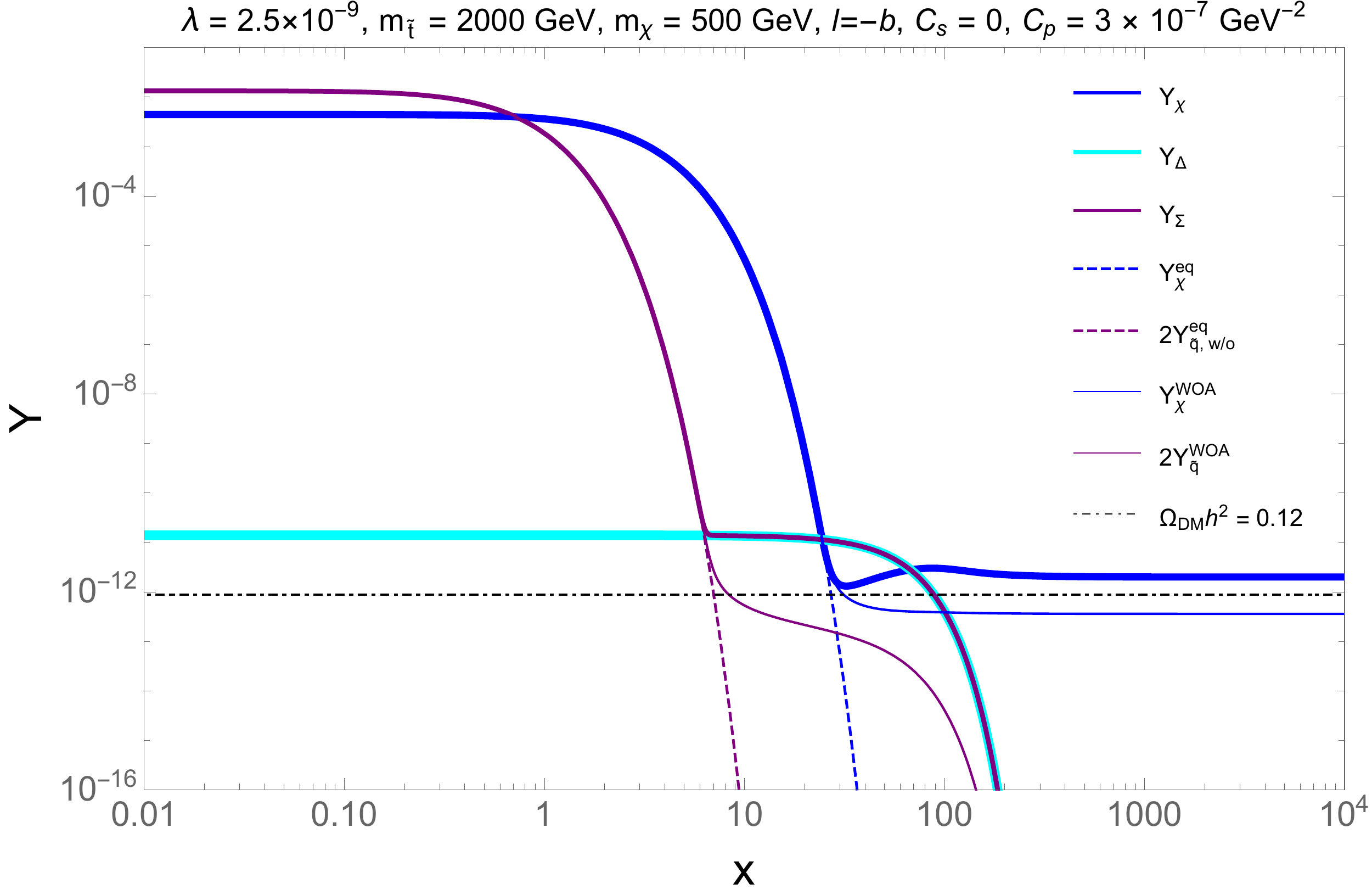}
\hfill
\includegraphics[width=0.48\textwidth]{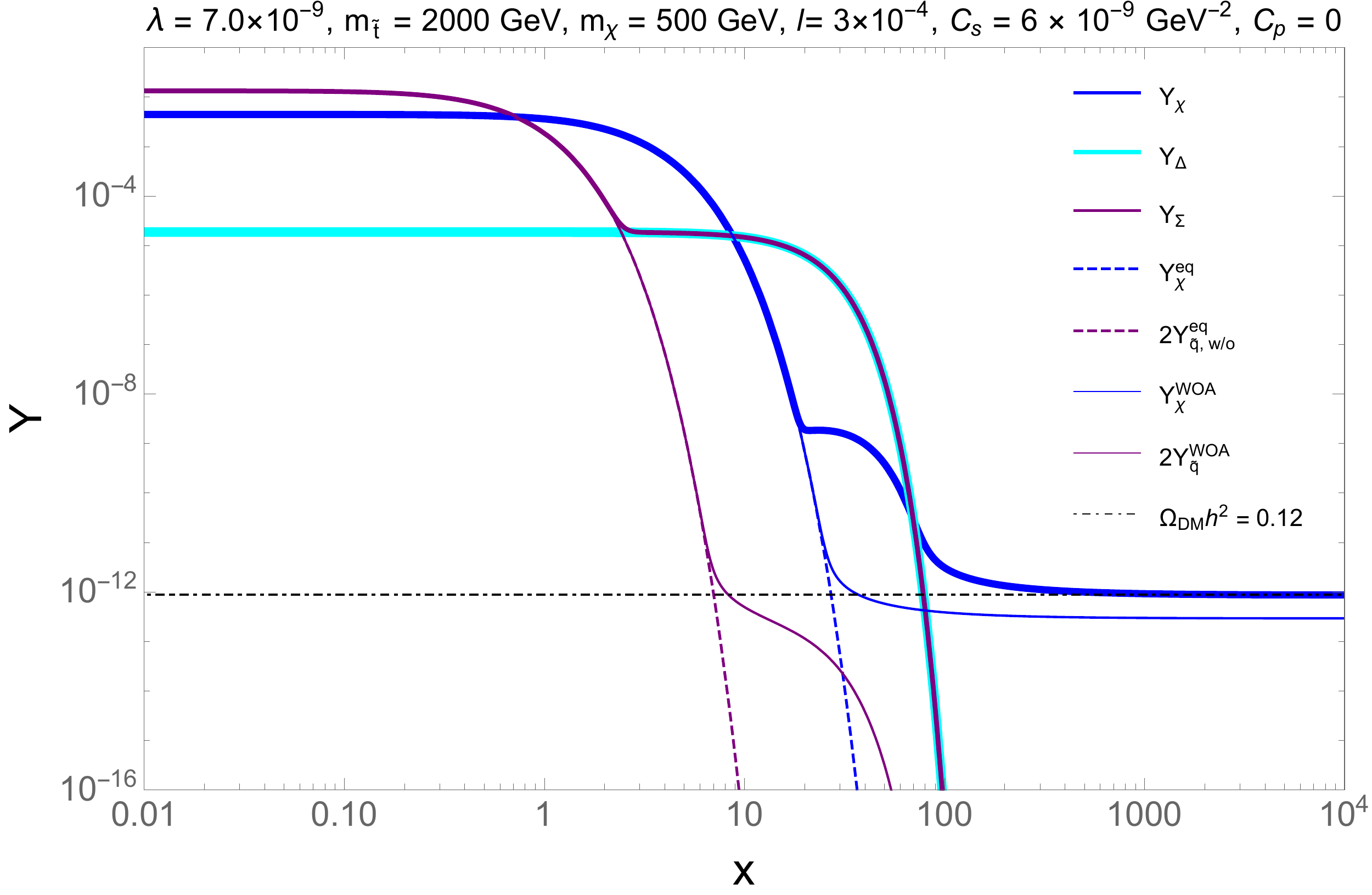}
\caption{
Left panel: evolution for the $p$-wave-dominated annihilation choice
$C_s=0$ and
$C_p=3\times10^{-7}~\mathrm{GeV}^{-2}$, with the remaining parameters
equal to those of Fig.~\ref{fig:benchmark_evolution}.
Right panel: evolution for the benchmark $s$-wave annihilation choice but
with the larger lepton asymmetry
$l=3\times10^{-4}$.
The comparison illustrates the effects of the late-time $\chi\chi$
annihilation rate and of the chemical-potential dependence of the
asymmetry-transfer process.
}
\label{fig:pwave_lepton}
\end{figure}

We compare the benchmark $s$-wave case,
\begin{equation}
C_s=6\times10^{-9}~\mathrm{GeV}^{-2},
\qquad
C_p=0,
\end{equation}
with the representative $p$-wave case shown in the left panel of
Fig.~\ref{fig:pwave_lepton},
\begin{equation}
C_s=0,
\qquad
C_p=3\times10^{-7}~\mathrm{GeV}^{-2}.
\end{equation}

The qualitative evolution in the $p$-wave case remains the same as in the
benchmark $s$-wave scenario: the symmetric partner population is first
depleted, a surviving asymmetry remains in $Y_\Delta$, and this asymmetry is
subsequently transferred to $\chi$.  The main difference is that the
$p$-wave annihilation cross section decreases with temperature,
\begin{equation}
\langle\sigma v\rangle_{\chi\chi}
\simeq
C_p\frac{T}{m_\chi},
\end{equation}
so that the annihilation of newly produced $\chi$ particles becomes less
efficient at late times.  Consequently, the post-transfer decrease of
$Y_\chi$ is milder and the final relic abundance is larger.

For the representative benchmark, the $p$-wave choice gives
$\Omega_{\rm DM}h^2\simeq0.27$, compared with
$\Omega_{\rm DM}h^2\simeq0.12$ for the $s$-wave choice.  This difference
does not signal a qualitatively different asymmetry-transfer mechanism.
Rather, it reflects the different annihilation efficiencies of the
injected $\chi$ population at the time of transfer.

\subsection{Sensitivity to the lepton asymmetry}
\label{subsec:lepton_asymmetry_dependence}

We finally examine the dependence of the results on the lepton asymmetry.
As discussed in Sec.~\ref{sec:chemical_potential}, we do not explicitly
model the sphaleron-induced redistribution of asymmetries. Instead, the
two choices of $l$ are used as representative inputs to assess the
sensitivity of the asymmetry-transfer mechanism to the lepton asymmetry.

The corresponding relic-density curves are shown in
Fig.~\ref{fig:relic_lambda_mchi}. The larger lepton asymmetry induces a
different quark chemical potential through the conservation equations,
which modifies the effective equilibrium abundances entering the
conversion terms and can therefore affect the detailed evolution of
$Y_\Delta$ and $Y_\chi$, as illustrated in the right panel of
Fig.~\ref{fig:pwave_lepton}. Nevertheless, the overall behavior remains
qualitatively unchanged: the symmetric partner population is depleted by
QCD annihilations, a surviving asymmetric component remains in the partner
sector, and this asymmetry is subsequently transferred to $\chi$.

For the larger lepton asymmetry, the magnitude of $\mu_q/T$ is enhanced
relative to the $l=-b$ benchmark, while remaining within the range where
the linearized treatment of the chemical potentials adopted in
Sec.~\ref{sec:chemical_potential} is applicable.  The resulting
relic-density curve can therefore differ quantitatively from the
$l=-b$ case, but it retains the same characteristic dependence on
$\lambda$ and the same qualitative transition between
asymmetry-transfer-dominated and thermally dominated relic abundance.

The two choices of $l$ should therefore be interpreted as a robustness test
rather than as two explicitly modeled cosmological histories. The
persistence of the same qualitative intermediate-coupling regime indicates
that the central asymmetry-transfer mechanism is not sensitive to the
precise value of the lepton asymmetry.

\subsection{Summary of the numerical picture}
\label{subsec:numerical_summary}

The numerical results can be summarized by the following physical picture.  The
partner sector initially contains both symmetric and asymmetric
populations.  
QCD annihilations efficiently deplete the symmetric component,
reducing $Y_\Sigma$ while leaving the net partner asymmetry
$Y_\Delta$ essentially unchanged.
At later times, the surviving asymmetric component is transferred to
$\chi$ through the portal interaction,
\begin{equation}
Y_\Delta
\;\longrightarrow\;
Y_\chi.
\end{equation}

The final dark matter abundance is consequently controlled by three
interrelated ingredients:
\begin{enumerate}
\item the amount of asymmetry that survives QCD depletion and early
conversion;
\item the time at which the surviving asymmetry is transferred to $\chi$;
and
\item the efficiency of $\chi\chi$ annihilation after the transfer.
\end{enumerate}

This explains the characteristic dependence on the portal coupling.
For large $\lambda$, the asymmetry is transferred too early, and the
resulting $\chi$ population is largely processed by annihilations before
it can contribute to the frozen dark matter abundance.  For very small
$\lambda$, the transfer occurs too late, after $\chi\chi$ annihilation has
become inefficient, resulting in an overabundance.  An intermediate range
of $\lambda$ allows the transfer to occur at a time when a substantial
fraction of the injected $\chi$ population can survive, making it possible
to reproduce the observed relic abundance.

The same picture also explains why no compressed mass spectrum is required.
The relevant reservoir is the asymmetric component of the partner
population, rather than its thermally suppressed symmetric abundance.
Consequently, hierarchical spectra with
$m_{\tilde q}\gg m_\chi$ can remain viable.

The comparison of the two representative lepton asymmetries shows that
the detailed chemical-potential evolution can modify the numerical results
without changing the basic mechanism.  Together with the $s$-wave/$p$-wave
comparison, this demonstrates that the central result is not tied to a
particular annihilation mode or to a particular choice of the lepton
asymmetry.

The numerical analysis therefore provides a consistent realization of the
conceptual picture introduced in Fig.~\ref{fig:conceptual_plot}: a
primordial baryon asymmetry is temporarily stored in the partner sector,
the symmetric partner population is removed by QCD annihilation, and the
surviving asymmetric component is subsequently transferred to the
Majorana dark matter particle, thereby supplying the dominant contribution
to its relic abundance.

The numerical results presented above focus on the stop-like partner
$\tilde q=\tilde t$.  The stop case has an additional kinematic feature
associated with the relatively large top-quark mass: the two-body decay
$\tilde t\to\chi t$ becomes kinematically forbidden for sufficiently
compressed spectra, leading to a change in the dominant conversion channels
as discussed above.
Apart from this kinematic feature, the asymmetry-transfer mechanism
and its qualitative dependence on the portal coupling and mass hierarchy
are common to other choices of the colored scalar partner.  We therefore
turn in the next section to the collider implications for both stop-like and sbottom-like partner
scenarios.

\section{LHC constraints from long-lived particle searches}
\label{sec:collider}
The small couplings considered in Sec.~\ref{sec:numerical_analysis} can also make $\tilde{q}$ long lived on detector scales. 
For the benchmark point discussed, $m_{\tilde q}=2000~\mathrm{GeV}$, $m_\chi=500~\mathrm{GeV}$, and $\lambda=2.5\times10^{-9}$, the lifetime of $\tilde{q}$ is about $3~\mathrm{ns}$, corresponding to a proper decay length of $c\tau_{\tilde q}\simeq 1~\mathrm{m}$. 
Such a lifetime lies in the range relevant for long-lived-particle searches at the LHC.  
After production at the LHC, a long-lived $\tilde{q}$ hadronizes into an $R$-hadron, whose charged states can give rise to tracks with anomalously large ionization energy loss \cite{ATLAS:2019gqq}. 
Such signatures are directly probed by searches for heavy long-lived charged particles at the LHC \cite{ATLAS:2022pib}. 
We therefore examine the corresponding constraints on both $\tilde{t}$ and $\tilde{b}$ scenarios considered in this work.

We follow the ATLAS full Run-2 large-$dE/dx$ search and the corresponding public reinterpretation prescription. 
The auxiliary efficiency tables and model-independent limits are taken from the HEPData record \cite{hepdata.127994}. 
Signal events for $pp\to\tilde{q}\tilde{q}^{*}$ are generated with \textsc{MadGraph5\_aMC@NLO} \cite{Alwall:2014hca}, followed by parton showering and hadronization with \textsc{Pythia}~8 \cite{Bierlich:2022pfr}. 
We use the NNLO+NNLL predictions for $\widetilde t$ and $\widetilde b$ pair-production cross sections from \textsc{NNLL-Fast} \cite{Beenakker:2016lwe}. 
The pair-production cross sections are taken to be the same for a given $m_{\tilde q}$, since $\tilde{q}\tilde{q}^{*}$ production is dominated by QCD interactions and the electroweak contributions are subdominant.
For the collider recast, we fix $m_\chi=300~\mathrm{GeV}$ as a representative non-compressed benchmark.  
A detailed treatment of compressed spectra, for which the event-level acceptance can depend on the decay kinematics, is not considered in the present analysis. 
The signal yields are compared with model-independent $95\%$ CL limits. 

\begin{figure}[htbp]
    \centering
    \includegraphics[width=\textwidth]{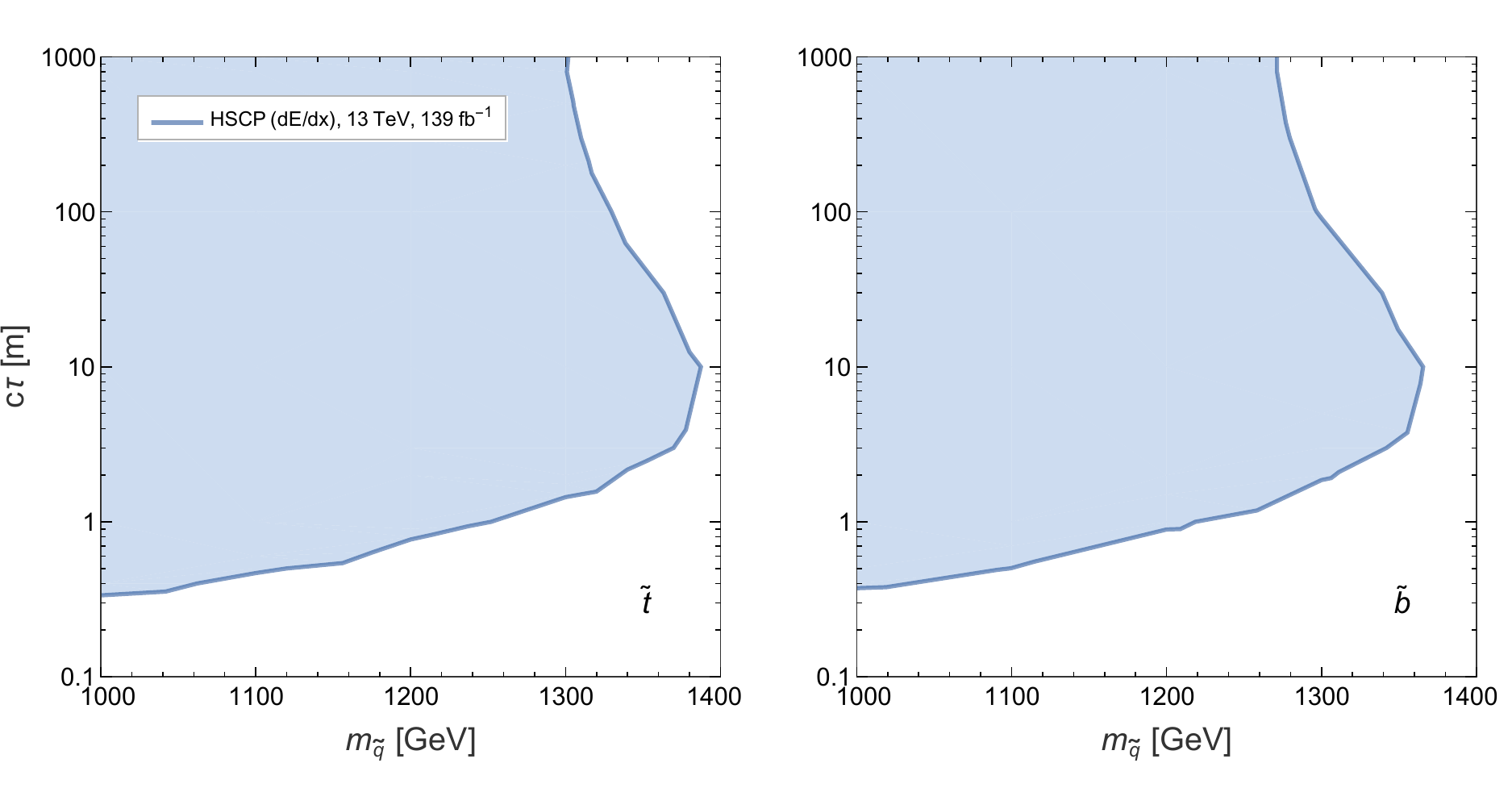}
    \caption{Recast of the observed $95\%$ CL exclusion regions from the ATLAS large $dE/dx$ HSCP search at $\sqrt{s}=13~\mathrm{TeV}$ with $139~\mathrm{fb}^{-1}$ in the $(m_{\widetilde q},c\tau)$ plane for the $\widetilde t$ (left) and $\widetilde b$ (right) scenarios, with $m_\chi=300~\mathrm{GeV}$. The shaded regions are excluded.}
    \label{fig:hscp_exclusion}
\end{figure}

The results are shown in Fig.~\ref{fig:hscp_exclusion}. 
For the non-compressed spectra considered in this recast, the proper decay length approximately scales as
\begin{equation}
c\tau_{\tilde q}\simeq
2~\mathrm{m}
\Big(\frac{\lambda}{2.5\times 10^{-9}}\Big)^{-2}
\Big(\frac{m_{\tilde q}}{1\ {\rm TeV}}\Big)^{-1}.
\end{equation}
The strongest limits are obtained when $c\tau \sim 10~\mathrm{m}$, for which $m_{\tilde{q}}$ up to about $1.4~\mathrm{TeV}$ are excluded. 
This reflects the balance between the loss of track efficiency at shorter lifetimes and the reduced $E_T^{\rm miss}$ trigger efficiency in the detector-stable limit. 
The slight difference between the $\tilde{t}$ and $\tilde{b}$ contours mainly originates from the different charge composition of their $R$-hadrons, leading to a somewhat larger signal acceptance in the $\tilde{t}$ case.

At shorter decay lengths, searches targeting displaced decays can provide complementary sensitivity to the HSCP constraints considered here~\cite{CMS:2020iwv,Heisig:2024xbh}. 
The cosmological analysis motivates $m_{\tilde q}$ values extending beyond the reach of the present HSCP recast. For such heavy states, searches combining ionization and time-of-flight information can improve the identification of slow $R$-hadrons and may further extend the sensitivity at the HL-LHC~\cite{Cerri:2018rkm,ATLAS:2025fdm}. 
A substantial extension into the multi-TeV mass range would ultimately require higher-energy hadron colliders~\cite{Cohen:2014hxa}.

\section{Conclusion}
\label{sec:conclusion}

We have studied a mechanism in which a primordial baryon asymmetry is temporarily stored in a heavier colored partner sector and is subsequently transferred to a self-conjugate dark matter particle. In the simplified model considered here, the dark matter particle is a Majorana fermion $\chi$, while the partner $\tilde q$ is a complex scalar carrying baryon number and coupled to a Standard Model quark through a small portal interaction. Unlike conventional asymmetric dark matter scenarios, the dark matter particle itself does not need to carry a conserved particle-antiparticle asymmetry. Instead, the partner sector acts as a temporary reservoir for the primordial asymmetry.

At early times, the partner population contains both symmetric and asymmetric components. QCD interactions efficiently deplete the symmetric component through $\tilde q\tilde q^*$ annihilation, while leaving the net asymmetry, 
$Y_\Delta=Y_{\tilde q}-Y_{\tilde q^*}$, 
essentially unchanged. For sufficiently small portal couplings, the conversion processes do not maintain chemical equilibrium throughout this stage, allowing a substantial asymmetric component to survive after the symmetric population has disappeared. This surviving asymmetry can subsequently be transferred to $\chi$ through decays and scattering processes.

We have incorporated the chemical-potential dependence of the conversion processes in a self-consistent way. The SM chemical potentials are determined from the baryon and lepton asymmetries, electric charge neutrality, and the chemical-equilibrium conditions enforced by rapid SM interactions, while the partner asymmetry $Y_\Delta$ is evolved directly with the Boltzmann equations. In particular, we do not impose $\mu_{\tilde q}=\mu_q$, since the small portal coupling need not maintain chemical equilibrium between the partner and the SM quark. 

The numerical results demonstrate that the final dark matter abundance is determined by the interplay of three effects: the amount of asymmetry surviving the depletion of the symmetric partner population, the time at which this surviving asymmetry is transferred to $\chi$, and the efficiency of $\chi\chi$ annihilation after the transfer. The portal coupling $\lambda$ plays a central role because it controls the timing of the transfer. For large $\lambda$, the asymmetry is transferred too early, when the produced $\chi$ particles can still efficiently annihilate, leaving a relic abundance close to the ordinary thermal result. For sufficiently small $\lambda$, the transfer occurs too late, after $\chi\chi$ annihilation has become inefficient, and the transferred component can instead lead to an overabundance. An intermediate range of portal couplings therefore allows the transferred asymmetry to provide a substantial, and potentially dominant, contribution to the observed dark matter abundance.

An important feature of this mechanism is that it does not require a compressed mass spectrum. We find viable examples with a substantial hierarchy between the partner and dark matter masses, including benchmark configurations with $m_{\tilde q}\gg m_\chi$. In such cases, the partner abundance relevant for the mechanism is the surviving asymmetric component rather than the thermally suppressed symmetric abundance. The dependence on the mass hierarchy can nevertheless be significant because changing $m_\chi$ modifies the relative timing of partner depletion, asymmetry transfer, and dark matter freeze-out. Near-degenerate spectra exhibit additional kinematic structure, particularly in the stop-like case, where the two-body decay $\tilde t\to\chi t$ can become forbidden and scattering or higher-body decay processes can dominate the conversion history.

The annihilation properties of $\chi$ provide another important source of variation. When the asymmetry is transferred after dark matter freeze-out, the final relic abundance depends sensitively on the annihilation rate at the time of injection. We therefore find a larger relic abundance for the representative $p$-wave benchmark than for the corresponding $s$-wave case, reflecting the suppression of the $p$-wave annihilation rate at late times. This dependence does not alter the underlying asymmetry-transfer mechanism, but determines how much of the transferred population survives as dark matter.

We have also examined the sensitivity to the lepton asymmetry by comparing $l=-b$ with a substantially larger representative value, $l=3\times10^{-4}$. The different lepton asymmetries modify the SM quark chemical potentials and consequently affect the detailed evolution of the conversion processes and the final relic density. Nevertheless, the qualitative behavior remains unchanged: QCD annihilation removes the symmetric partner population, a residual asymmetric component survives, and this component is transferred to $\chi$ at later times. This indicates that the basic mechanism is not tied to a particular choice of the lepton asymmetry, although a quantitative treatment of the sphaleron-active era would be required for a more complete description of the chemical-potential evolution.

Finally, the small portal couplings relevant for late asymmetry transfer can make the colored partner long lived on detector scales. For the benchmark point with $m_{\tilde t}=2~\mathrm{TeV}$, $m_\chi=500~\mathrm{GeV}$, and $\lambda=2.5\times10^{-9}$, the predicted lifetime is of order a few nanoseconds, corresponding to a proper decay length of order one meter. Such states can hadronize into long-lived $R$-hadrons and are therefore subject to searches for heavy stable or metastable charged particles. Our recast of the ATLAS Run-2 large-$dE/dx$ search shows that existing HSCP searches constrain part of the parameter space for both stop-like and sbottom-like partners, while heavier partner masses, such as those favored by the cosmological mechanism, remain beyond current limits and constitute a target for future higher-luminosity or higher-energy collider searches.

Overall, our results demonstrate that a heavier colored partner can serve as a temporary reservoir for a primordial baryon asymmetry and subsequently transfer this asymmetry to a self-conjugate dark matter particle. The mechanism operates through the interplay of QCD depletion, delayed portal conversion, and residual dark matter annihilation, and can reproduce the observed relic abundance without requiring the dark matter particle itself to carry a conserved asymmetry or requiring a compressed spectrum. This framework thereby establishes a qualitatively distinct pathway from conventional asymmetric dark matter, in which a primordial baryon asymmetry can manifest as the observed dark matter density even when the dark matter particle is its own antiparticle. The simultaneous possibility of late cosmological conversion and detector-scale long-lived colored particles provides a direct connection between the relic-density mechanism and long-lived-particle searches at colliders.

\section*{Acknowledgments}
F.L. and M.T. would like to thank S. Matsumoto, S. Shirai and T.T. Yanagida for valuable discussions during their stay and visit at the Kavli IPMU. F.L. is supported by the Sun Yat-sen University Science Foundation. M.T. is supported by the Fundamental Research Funds for the Central Universities, the One Hundred Talent Program of Sun Yat-sen University, China, and the Guangdong Natural Science Foundation (Project No. 2026A1515012641).


\bibliographystyle{bibstyle_jhep}
\bibliography{ref_hdlt}
\end{document}